\documentclass[runningheads]{llncs}
\usepackage{graphicx}

\usepackage{tikz}
\usepackage{comment}
\usepackage{amsmath,amssymb} %
\usepackage{color}

\usepackage{booktabs}
\usepackage[utf8]{inputenc} %
\usepackage[T1]{fontenc}    %
\usepackage{lmodern}
\usepackage{lmodern}  %
\usepackage{url}%
\usepackage{booktabs}       %
\usepackage{amsfonts}       %
\usepackage{nicefrac}       %
\usepackage{microtype}      %
\usepackage{xcolor} %
\usepackage{amsmath}%
\usepackage{bm}
\usepackage{multirow}
\usepackage{subcaption}
\usepackage{multicol}
\usepackage{cite}  %

\usepackage{hyperref}

\def\R{\mathbb{R}}
\DeclareMathSymbol{:}{\mathord}{operators}{"3A}
\newcommand{\powerset}{\raisebox{.15\baselineskip}{\Large\ensuremath{\wp}}}

\begin{document}
\pagestyle{headings}
\mainmatter
\def\ECCVSubNumber{7301}  %

\title{AudioScopeV2: Audio-Visual Attention Architectures for Calibrated Open-Domain On-Screen Sound Separation} %

\titlerunning{AudioScopeV2: Audio-Visual Attention for Calibrated On-Screen Separation}
\author{Efthymios Tzinis\inst{1,2}\thanks{Work done during an internship at Google Research.}\and
Scott Wisdom\inst{1} \and
Tal Remez\inst{1}\and
John R.\ Hershey\inst{1}}
\authorrunning{E. Tzinis et al.}
\institute{
Google Research
\and
University of Illinois Urbana-Champaign
\\
\email{etzinis2@illinois.edu}, \email{\{scottwisdom,johnhershey\}@google.com}}
\maketitle

\begin{abstract}
We introduce AudioScopeV2, a state-of-the-art universal audio-visual on-screen sound separation system which is capable of learning to separate sounds and associate them with on-screen objects by looking at in-the-wild videos. We identify several limitations of previous work on audio-visual on-screen sound separation, including the coarse resolution of spatio-temporal attention, poor convergence of the audio separation model, limited variety in training and evaluation data, and failure to account for the trade off between preservation of on-screen sounds and suppression of off-screen sounds. We provide solutions to all of these issues. Our proposed cross-modal and self-attention network architectures capture audio-visual dependencies at a finer resolution over time, and we also propose efficient separable variants that are capable of scaling to longer videos without sacrificing much performance. We also find that pre-training the separation model only on audio greatly improves results.  For training and evaluation, we collected new human annotations of on-screen sounds from a large database of in-the-wild videos (YFCC100M). This new dataset is more diverse and challenging. Finally, we propose a calibration procedure that allows exact tuning of on-screen reconstruction versus off-screen suppression, which greatly simplifies comparing performance between models with different operating points. Overall, our experimental results show marked improvements in on-screen separation performance under much more general conditions than previous methods with minimal additional computational complexity.
\end{abstract}

\section{Introduction}

\begin{figure}[htb!]
    \centering
    \begin{subfigure}[h]{0.638\linewidth}
      \includegraphics[width=\linewidth]{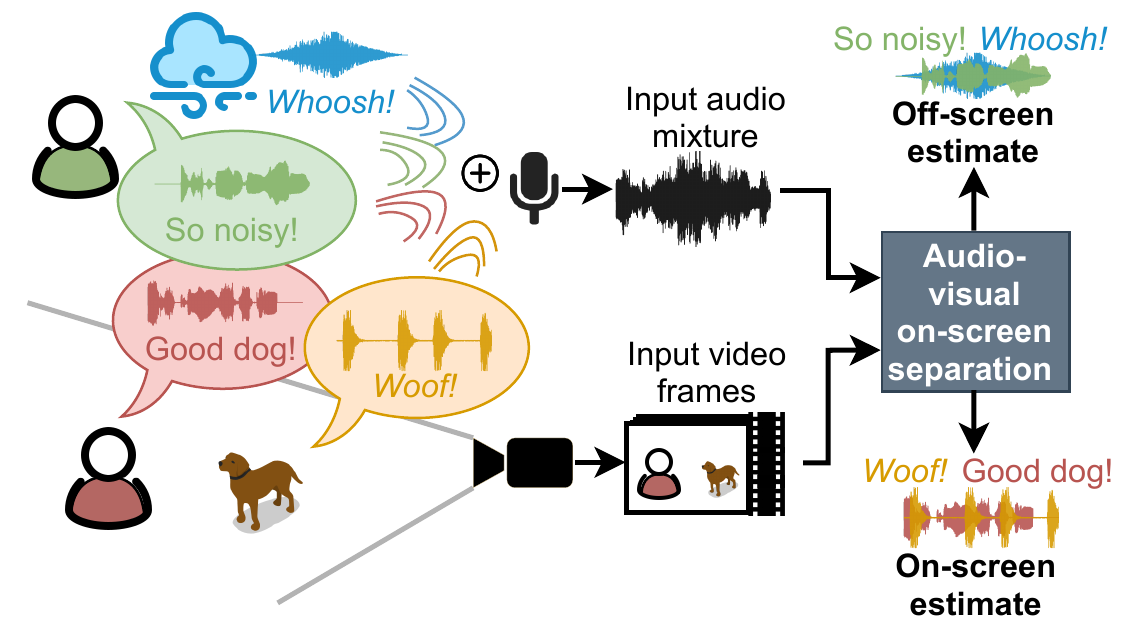}
      \caption{General audio-visual on-screen sound separation task.}
      \label{fig:intro:task} 
     \end{subfigure}
     \vline
     \begin{subfigure}[h]{0.353\linewidth}
     \centering
      \includegraphics[width=\linewidth]{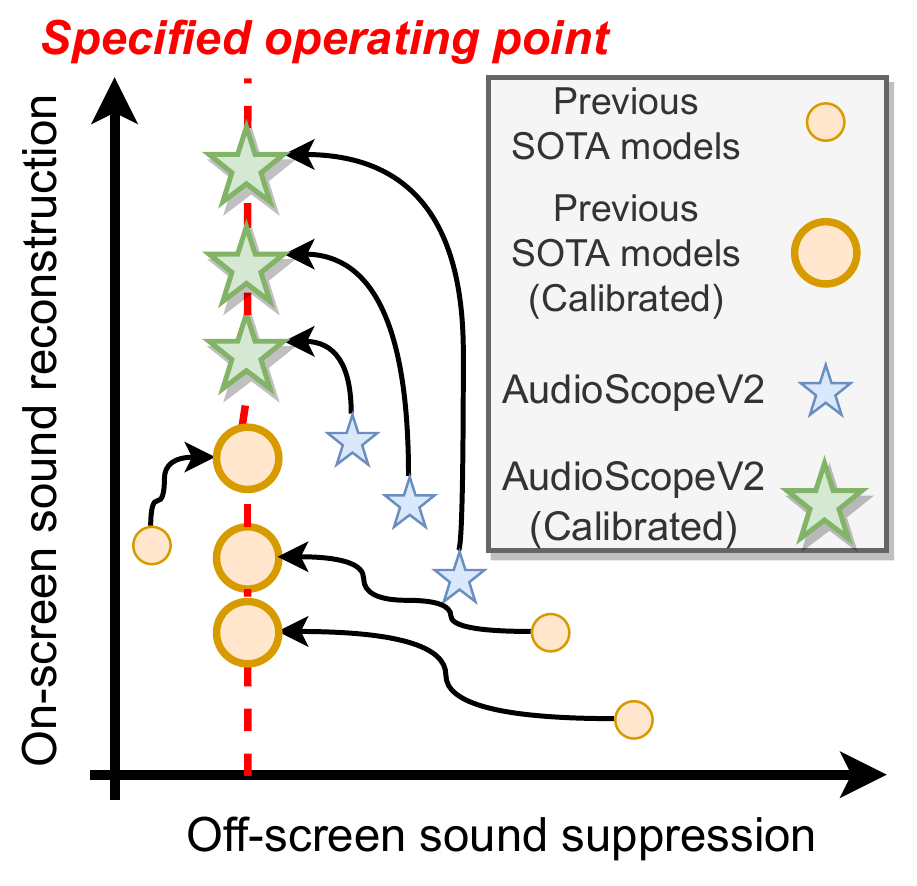}
      \caption{Proposed calibration.}
      \label{fig:intro:motivation} 
     \end{subfigure}
    \caption{Left: illustration of the task where our models make no assumptions about the existence, type, or count (up to a maximum) of on-screen and off-screen sources. Right: illustration of our proposed calibration procedure for on-screen separation models, which allows setting a specified tolerance level for off-screen sound suppression which allows more user control and easier model comparison.
    }
    \label{fig:intro}
\end{figure} 

Humans are able to effortlessly perceive sounds in a noisy scene, and associate them with any corresponding visible objects.
In audio processing, a corresponding challenge is to isolate sound sources from a mixture waveform and identify the associated visual appearance of each sound source. In this paper, we target the task of \emph{universal audio-visual on-screen sound separation}, where the goal is to recover only the sounds that originate from on-screen objects, regardless of the types of on-screen and off-screen objects as illustrated in Figure \ref{fig:intro:task}.

This is a difficult task for numerous reasons. In stark contrast to visual objects that generally occupy distinct regions of pixels, sound sources are superimposed in the time domain.
This imposes a challenge for unsupervised learning of audio-visual separation, because unlike their visual counterparts, the audio sources in a scene cannot be easily selected and aligned with video objects.  
Therefore, separating the constituent sources is needed, by conditioning separation on selected video objects and/or separating the audio \emph{a priori} before associating the sounds with video objects.   The \emph{a priori} separation of the sounds, which we pursue here, has a few advantages. Thanks to recent work \cite{MixITNeurIPS}, it can be learned in an unsupervised way, and it can handle an unknown number of sounds, including those that do not appear on-screen. Also, the individual separated sounds are available to downstream processes, in addition to the on-screen estimate.

Despite remarkable progress in the field of on-screen sound separation, most of these works are constrained to isolating only a specific set of sound classes that can appear on-screen such as speech \cite{ephrat2018looking,gao2021visualvoiceSpeechSeparationCrossModalConsistency,afouras2018conversationDeepAVSpeechEnhancement} or music \cite{gan2020musicGestureforAVSeparation ,gao2018learningToSeparateObjectFromUnlabeledVideo}. Although this strategy works well under a restricted domain, where such labeled data are available, the reliance on human labels precludes scaling to large open-domain data. Recent works have started to expand beyond music and speech to a wider variety of classes, such as using visual scene graphs to model audio-visual relationships \cite{chatterjee2021visual},
but this approach still requires labeled data to train a supervised object detector. Although the seminal works in on-screen sound separation proposed models that were somewhat invariant to the types of sources \cite{owens2018audioFirstOnScreen,gao2018learningToSeparateObjectFromUnlabeledVideo}, those systems were unable to be trained with real world videos mainly because they needed labeled videos in which the sources always appeared on-screen during training.  

Recently, AudioScope \cite{tzinis2021into} addressed several of the aforementioned problems using \emph{mixture invariant training} (MixIT) \cite{MixITNeurIPS} with synthetic mixtures of video soundtracks. 
The derived sources-to-soundtracks assignments from MixIT were used as pseudo-targets to self-supervise the training of an audio-visual sound separation model from in-the-wild videos, without requiring object detection modules or assuming that all sources have to be on-screen. 
However, AudioScope still suffers from generalization issues since
it relies on training data filtered by an unsupervised audio-visual coincidence model \cite{jansen2020coincidence}, which limits its generalization ability, as we show in our experiments. We also hypothesize that AudioScope's performance is limited by the simplicity of its visually guided spatio-temporal attention layer \cite{BahdanauCB14}, and the low temporal resolution (one frame per second) of its visual model. These factors may prevent AudioScope from capturing synchronization features which can be crucial for detecting the audio-visual interplay \cite{hershey2000audio,korbar2018cooperativeSelfSupervisedSynch,afouras2020selfsupAVObjectsOpticalFlow}.
Another limitation of AudioScope is the lack of ability to trade off between reconstruction of on-screen sounds and suppression of off-screen sounds. These models achieve an arbitrary operating point during training, which makes comparing performance between different models difficult (see Figure \ref{fig:intro:motivation}).

We propose solutions for all of the aforementioned problems and limitations:
\begin{enumerate}
    \item AudioScopeV2 leverages richer cross-modal and self-attention network architectures that capture audio-visual dependencies at a finer time resolution, as well as efficient separable variants that are capable of scaling to longer videos without sacrificing much performance (Section \ref{sec:model}). We also find that pre-training the audio separation model using MixIT greatly improves results.
    \item We provide a new dataset, for which we collected new human annotations of on-screen sounds from a large database of in-the-wild videos (YFCC100M \cite{thomee2016yfcc100m}), described in Section \ref{ssec:data_prep}. We show that our new proposed models both generalize and perform better on the more diverse and challenging evaluations sets compared to previous state-of-the-art methods.
    \item We propose a novel calibration procedure (Section \ref{ssec:metrics_and_calibration}) that allows precise tuning of on-screen reconstruction versus off-screen suppression, that can also be used to greatly simplify model comparison across different models that each have their own operating point.
\end{enumerate}
Dataset recipes, demos, and other supplementary material are available online \noindent\url{google-research.github.io/sound-separation/papers/audioscope-v2}.

\section{Relation to Prior Work}
Joint perception of audio and video modalities is not trivial, in part due to the problems of alignment between corresponding representations in each modality.
Nevertheless, a variety of works have shown promising results using multi-modal neural network architectures \cite{cheng2020lookListenAttendSSLAVRepresentationLearning,wu2019dualAttentionForAVEventLocalization,bertasius2021divisibleAttention,afouras2020selfsupAVObjectsOpticalFlow ,gao2019coSeparation,tzinis2021into,chatterjee2021visual}.  Audio-visual sound separation \cite{hershey2002audio,ephrat2018looking,afouras2018conversationDeepAVSpeechEnhancement}, and specifically separation of on-screen versus off-screen sounds \cite{owens2018audioFirstOnScreen}, has enjoyed remarkable performance improvements since the initial works. 
Important innovations have included using localization of objects \cite{wu2019dualAttentionForAVEventLocalization, zhu2021visually_guided_SS_localization,  lin2021unsupervisedSoundLocalizationIterativeContrastive}, forcing consistency between audio and visual representations \cite{lee2021CrossModalAffinityAudVisSpeechSeparation, gao2021visualvoiceSpeechSeparationCrossModalConsistency, arandjelovic2018objectsThatSound, gao2019coSeparation}, weakly-supervised \cite{rahman2021weaklySupervisedAVSoundSourceSeparation} and self-supervised \cite{rouditchenko2019selfAVCo_Segmentation, korbar2018cooperativeSelfSupervisedSynch, afouras2020selfsupAVObjectsOpticalFlow, tzinis2021into} approaches.

Recent work has shown that it is possible to train an open-domain audio-only universal sound separation model using a mask-based convolutional architecture regardless of the category of the sound \cite{kavalerov2019universal, tzinis2020two}. 
A related  direction is to extract sources of interest by conditioning separation networks using identity or multi-modal cues.  This has yielded performance improvements for speech \cite{wang2019voicefilter} as well as universal source separation \cite{tzinis2020improving,gfeller2021one,ochiai2020listenToWhatYouWant,kilgour2022text,liu2022separate}. 
However, these experiments relied on having sufficient supervised training data and were evaluated only on test sets with similar environmental conditions and sound distributions.  In order to extend the reach of this approach, methods have been proposed to train separation models with no access to ground truth clean sources by utilizing weak class labels \cite{pishdadian2020finding}, the spatial separability of the sources \cite{tzinis2019unsupervised,seetharaman2019bootstrapping,drude2019unsupervised} and self-supervision in the form of MixIT \cite{MixITNeurIPS}. These methods make it possible to learn separation of signals well outside the domains for which isolated source databases exist.  

An open question in audio-visual correspondence models concerns the level of processing at which audio and video objects can be aligned.  Typically audio-visual models have used high-level features at the output of neural networks to estimate correspondence between audio and video signals \cite{lee2021CrossModalAffinityAudVisSpeechSeparation,tzinis2021into,jansen2020coincidence,tian2020unified_multisensory_perception,korbar2018cooperativeSelfSupervisedSynch,chatterjee2021visual}. Such high-level representations may tend to focus on semantic information about the class of objects and sounds, especially when the features are computed at low video frame rates.  Such methods may work well for single instances of a class of object or sound, but may struggle with identification for multiple instances of a class, or for classes not seen during training.   
In contrast, there may be significant information in the correspondence between lower-level features. Mutual information between low-level features was used for audio-visual localization \cite{hershey2000audio}, and several more recent works have shown promising results for self-supervised audio-visual learning using low-level motion \cite{zhu2021visually_guided_SS_localization} and optical flow \cite{afouras2020selfsupAVObjectsOpticalFlow} features. Such features may help with generalization and instance-level correspondence by detecting synchronous dynamics of the audio and video, regardless of their semantic class.

Attention mechanisms can align representations across modalities, both at the level of semantic association and in terms of low-level correspondences.   %
An attention-based framework was recently used to modulate audio representations using motion-based visual features \cite{zhu2021visually_guided_SS_localization} for separation and localization. Conversely, modulating video features based on audio embeddings has also been used for speech separation \cite{lee2021CrossModalAffinityAudVisSpeechSeparation} as well as in AudioScope's spatio-temporal attention module \cite{tzinis2021into}. Other works combined self-attention layers \cite{vaswani2017attention} for modeling inter-modality temporal patterns, as well as cross-modal attention modules for intra-modality associations \cite{yu2021mpnSA_CMA_EventLocalization, cheng2020lookListenAttendSSLAVRepresentationLearning, wu2019dualAttentionForAVEventLocalization}, for sound localization and representation learning. One issue with self-attention is that its complexity grows quadratically with the dimensionality of the input length. We therefore propose a separable variants of our proposed architectures that factorize attention across different dimensions and modalities.  This strategy allows us to achieve similar performance to full self-attention with a much lower computational footprint.  Other separable attention mechanisms have emerged recently \cite{bertasius2021divisibleAttention, li2021vidtrSeparableAttentionVideo}, but our approach differs in that we process and capture intra-modality patterns from both audio and video features.

\begin{figure}[ht!]
  \centering
  \includegraphics[width=0.75\linewidth]{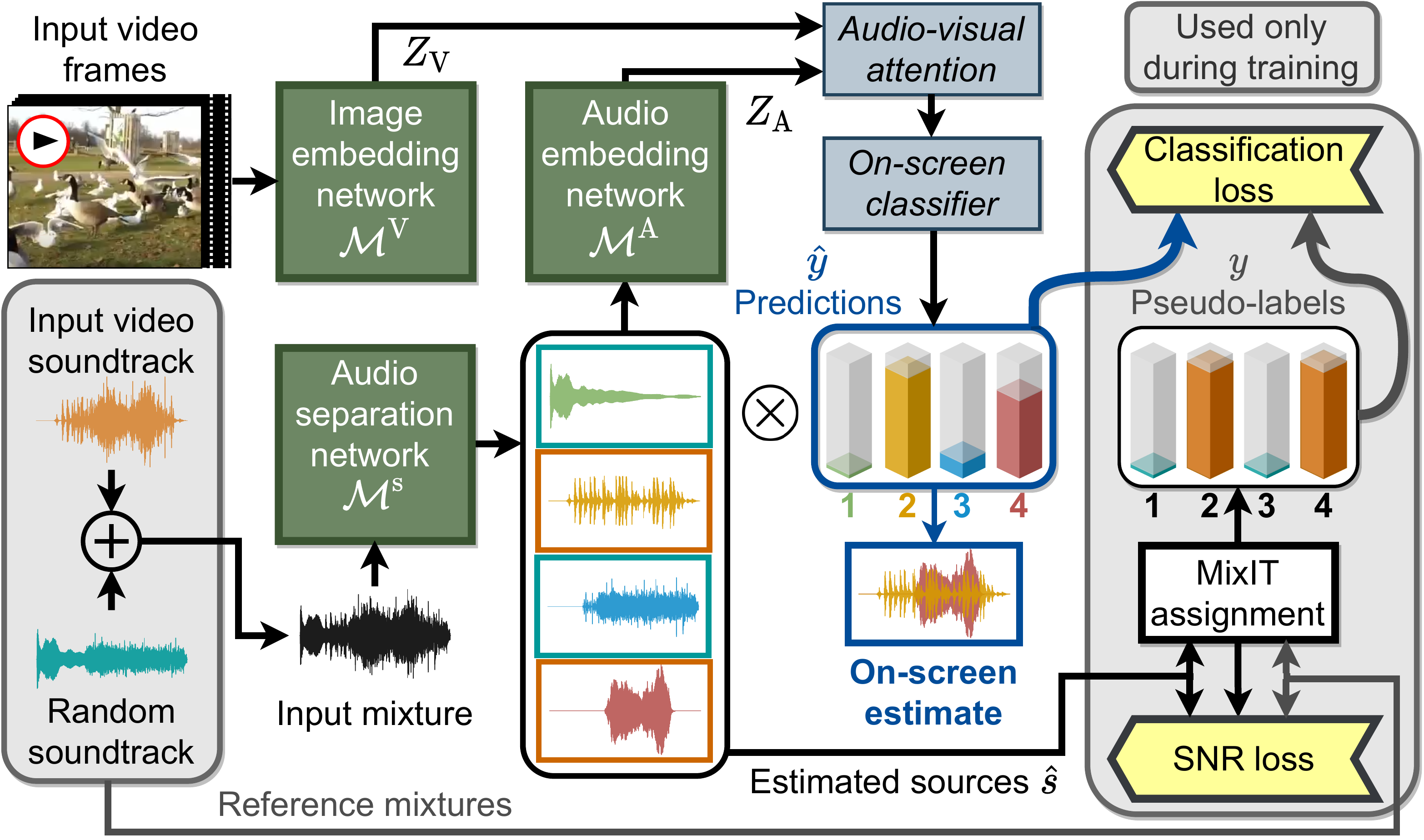}
\caption{Illustration of AudioScopeV2's architecture and its training procedure.
}
  \label{fig:schematic}
\end{figure}

\section{Model Architecture}
\label{sec:model}

Figure \ref{fig:schematic} illustrates the architecture and training procedure of AudioScopeV2.  The run-time architecture takes as input the video frames and a mixture waveform $x$. The audio separation module estimates $M=4$ sources. An audio embedding network is run on each of the estimated sources, producing audio embeddings $Z_\mathrm{A}$. In parallel, an image embedding network processes each input video frame independently and produces the visual embeddings $Z_\mathrm{V}$. The audio and video embeddings are fed to an audio-visual attention network, for which we propose a family of attention-based architectures. The output of the audio-visual attention network is passed to a final on-screen classifier head which produces a probability $\hat{y}_m$ corresponding to the event that the source $m$ originates from an on-screen object. Finally, the probabilities $\hat{y}_m$ are used as weights to mix the separated sources $\hat{s}_m$ together, producing an estimate of on-screen audio. The training procedure for this model is also illustrated, where we create mixtures of mixtures (MoMs) for audio by adding the soundtrack from another random video. A MixIT SNR loss is computed by finding the best combination of estimated sources to approximate each one of the reference mixtures in terms of SNR. The assignments of these best combinations are used as pseudo-labels in the classification loss. We describe each of these components in more detail below.
\subsection{Separation module}
\label{model:separation}
The separation module $\mathcal{M}^\mathrm{S}$ uses a dilated convolutional architecture \cite{kavalerov2019universal} with learnable encoder and decoder, which was also used in \cite{tzinis2021into}.  This module takes as input a mixture waveform $x \in \R^{T'}$, estimates $M$ masks in the encoded latent space, and outputs $M$ estimated source waveforms $\hat{s} \in \mathbb{R}^{M \times T'}$. The latter are forced to add up to the input mixture through a mixture consistency layer \cite{wisdom2018consistency}.  
However, in contrast to AudioScope \cite{tzinis2021into}, our separation model is not conditioned on global visual embeddings, for two reasons. First, visual conditioning was not shown to be effective \cite{tzinis2021into}, and second, this allows us to use MixIT to pre-train the separation module on all YFCC100M \cite{thomee2016yfcc100m} audio tracks to provide a better initialization for training the audio-visual model. Thus, AudioScope$^*$ refers to our AudioScope implementation with the aforementioned source separation module.

\subsection{Audio and video embedding networks}
\label{model:embedding_networks}
Features are extracted for the $M$ estimated sources $\hat{s}$ and the corresponding $T$ input video frames ($128 \times 128$ pixels), using a MobileNetV1 architecture~\cite{howard2017mobilenets}, as in \cite{tzinis2021into}. The audio encoder takes as input the log mel-scale spectrogram of the separated source waveforms $\hat{s}$, and audio features are extracted from the $23$rd layer. The visual embedding network is applied to each of the $T$ input video frames independently and extracts features with $8 \times 8$ spatial locations. Audio and video features are converted to a common depth $D=128$ with a dense layer.

\subsection{Audio-visual attention}
\label{model:attention}
We propose attention mechanisms to identify dependencies across the $M$ estimated audio sources, space, and time between the audio features $Z_\mathrm{A} \in \R^{M \times T \times D}$ and the video features $Z_\mathrm{V} \in \R^{G \times T \times D}$. Our video encoder provides $G=8^2$ spatial locations, and the time dimension $T$ is shared across both tensors.  Specifically, we propose to use \emph{audio-visual self-attention} (SA) \cite{vaswani2017attention}, which treats the audio sources and visual locations as a joint attention space (see Section \ref{model:attention:sa}),  and \emph{cross-modal attention} (CMA) layers, which perform attention between audio sources and visual locations, but avoid uni-modal attention between sources and between spatial locations (see Section \ref{model:attention:cma}). For both the SA and CMA attention we consider two settings: a \emph{joint attention} setting, in which attention operates jointly over time and space / sources, and a \emph{separable attention} setting, in which attention across time is interleaved with attention across spatial locations and sources. The joint attention scales quadratically with the product of dimensions from all of the axes which is computed over whereas the more efficient separable variation factorizes the operation across each axes individually,
which makes an important difference in practice (see Section \ref{sec:results:efficiency}).

In the following formulations we use a slightly more general version of an attention layer \cite{BahdanauCB14} to show how attention operates across different axes of the corresponding tensors. Attention computes similarities between a packed tensor of queries $Q \in \R^{X_Q \times T_Q \times D}$ w.r.t. some packed keys $K \in \R^{X_K \times T_K \times D}$, where $D$ is the depth dimensionality of the tensors. The similarities are computed using a generalized version of the typical tensor inner product $\langle Z_1, Z_2 \rangle_{\mathcal{A}}$, which reduces across the specified dimensions $\mathcal{A}$ of the second tensor $Z_2$. Note that we assume that $Z_1$'s dimensions are a subset of $Z_2$'s. By using a scaled tensor inner-product \cite{vaswani2017attention} and a $\operatorname{softmax}_\mathcal{A}$ activation that averages over the dimensions specified by $\mathcal{A}$ at the output of the tensor product $\langle K, Q\rangle_{\{\mathrm{D}\}}$, we produce the resulting similarity tensor which modulates the values $V \in \R^{X_V \times T_V \times D}$:
\begin{equation}
    \begin{aligned}
\operatorname{Att}_{\mathcal{A}}(Q, K, V) = 
\langle \alpha, f_{\mathrm{V}}(V)\rangle_{\mathcal{A}}, 
\enskip
\alpha = \operatorname{softmax}_\mathcal{A}\left(\frac{1}{\sqrt{D}} 
\langle f_{\mathrm{K}}(K), f_{\mathrm{Q}}(Q) \rangle_{\{\mathrm{D}\}}
\right),
    \end{aligned}
    \label{eq:general_attention}
\end{equation}
where $Q$, $K$, $V$, and $\alpha$ are the query tensor, the key tensor, the value tensor, and the attention weight distribution tensor across the set of specified axes $\mathcal{A}$ of the value/key tensors.
For example, for an input query $Q$ of shape $X_Q\times T_Q \times D$ and value $V$ of shape $X_V\times T_V \times D$, $\operatorname{Att}_{\{\mathrm{X_V}, \mathrm{T_V}\}}(Q, V, V)$ performs attention over the first and second axes of $V$, yielding an output tensor of shape $X_Q\times T_Q \times D$. The dense layers $f_{\mathrm{Q}}$,  $f_\mathrm{V}$, $f_{\mathrm{K}}$ are trainable and applied to the depth dimension $D$.

\begin{figure}[t]
    \centering
    \begin{subfigure}[h]{0.28\linewidth}
      \includegraphics[height=8cm]{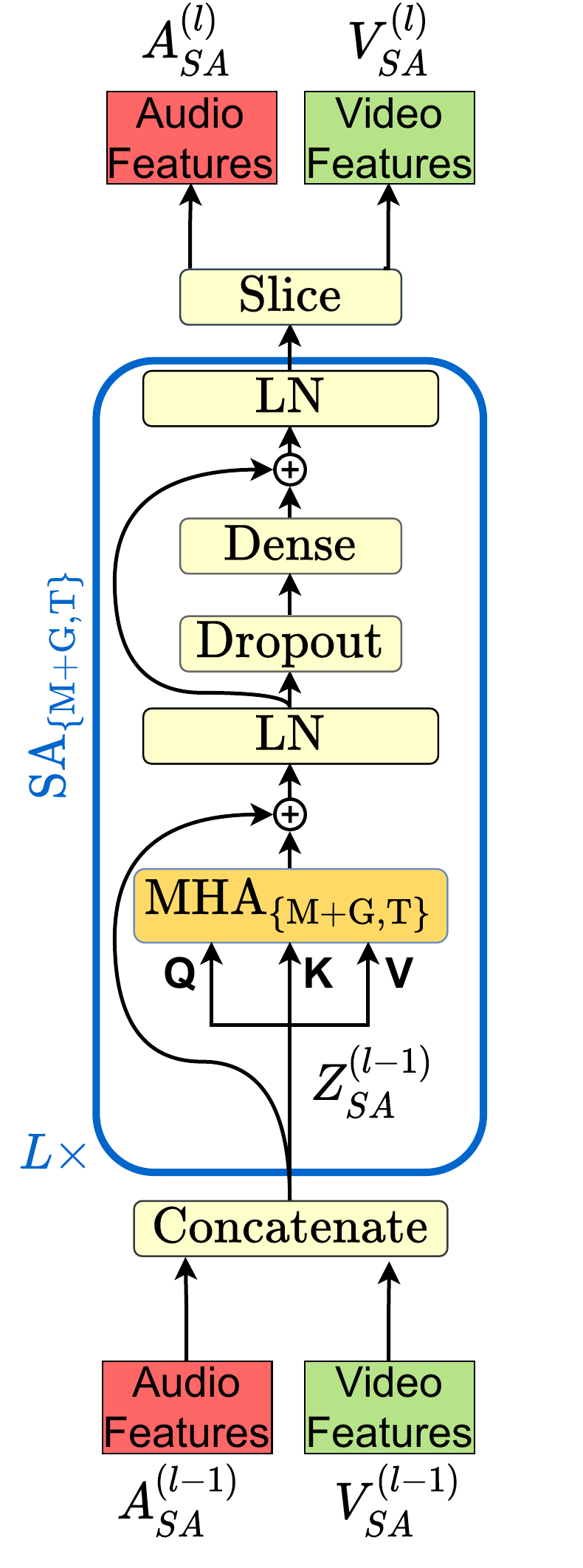}
      \caption{Self-Attention (SA).}
      \label{fig:sa_block} 
     \end{subfigure}
     \begin{subfigure}[h]{0.42\linewidth}
     \centering
      \includegraphics[height=8cm,trim={5pt 0 0 0},clip]{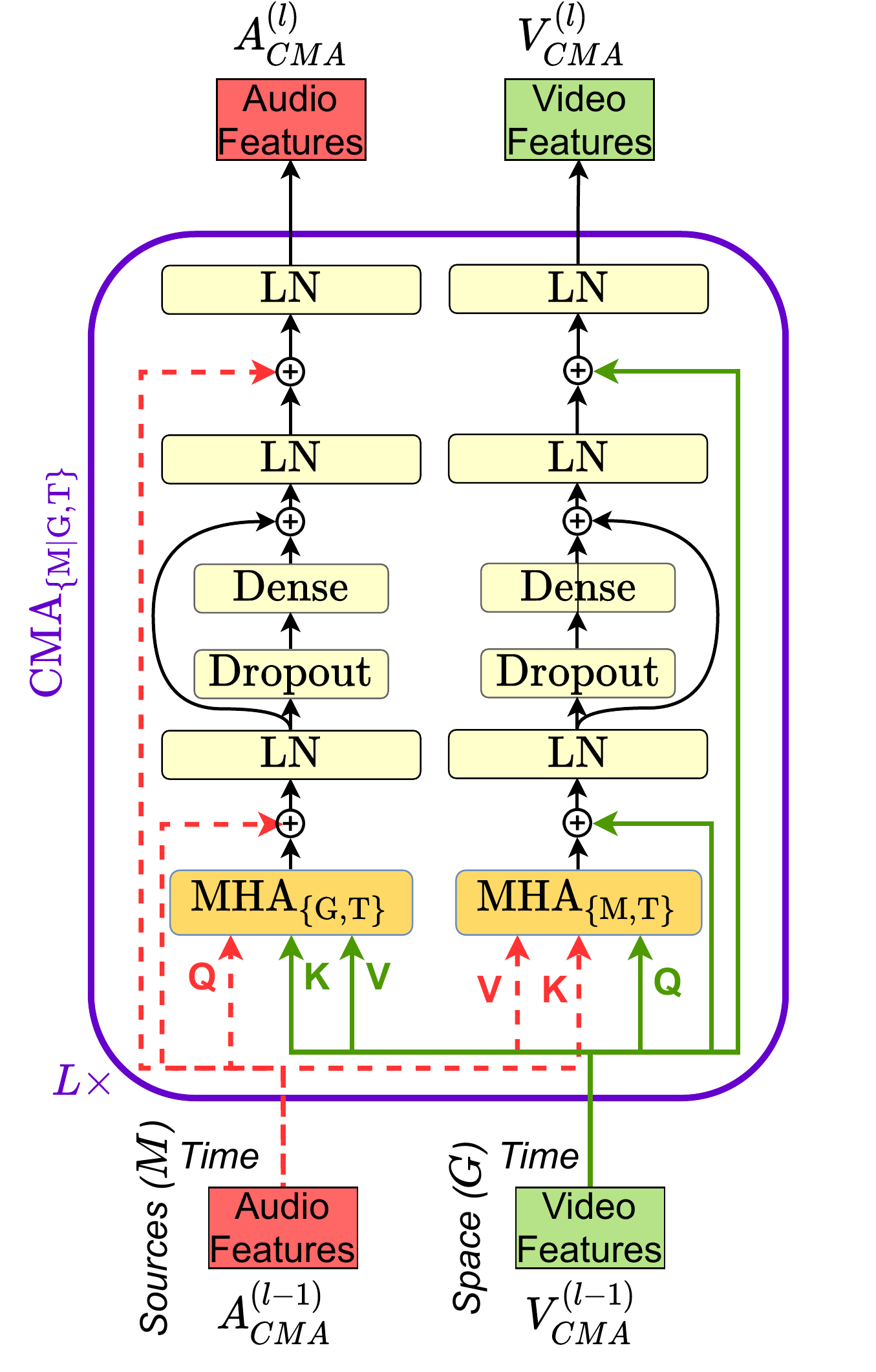}
      \caption{Cross-modal attention (CMA).}
      \label{fig:cma_block} 
     \end{subfigure}
  \begin{subfigure}[h]{0.2623\linewidth}
      \includegraphics[height=8cm]{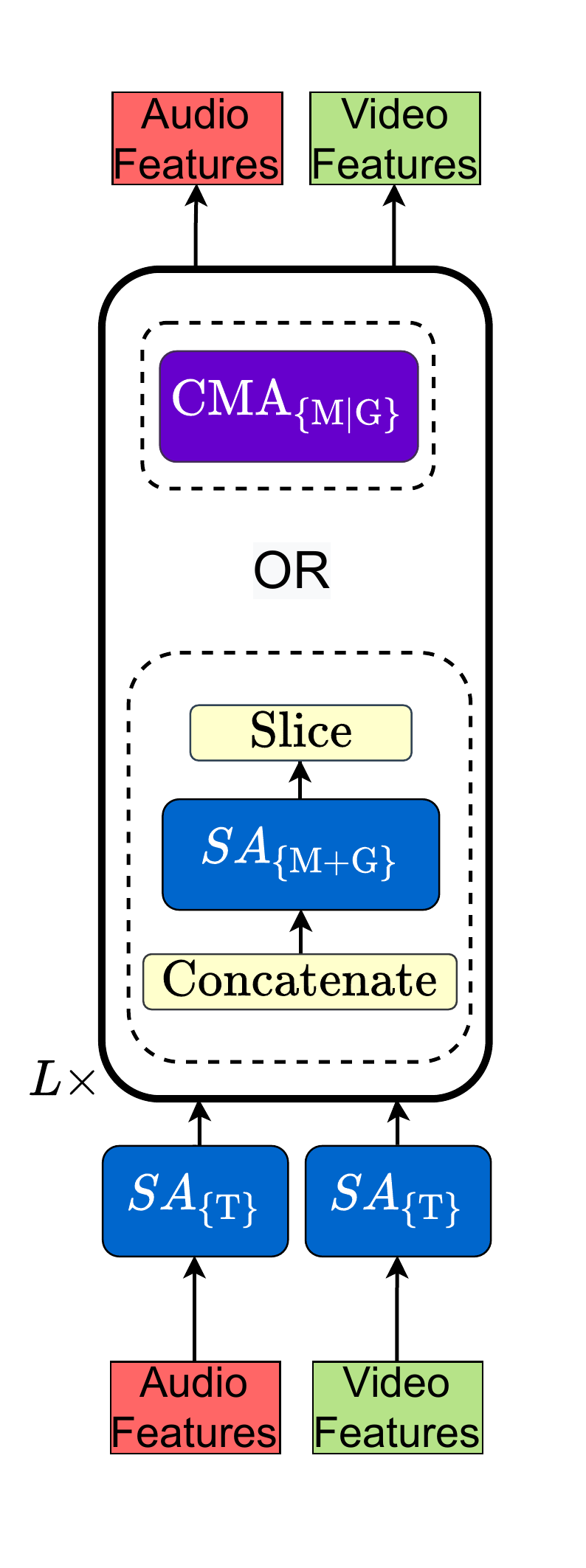}  
      \caption{Separable.}
      \label{fig:separable_cma_block} 
     \end{subfigure}\\
    \caption{AudioScopeV2's attention architectures for audio-visual alignment and feature extraction. Tensors are depicted by omitting batch and depth dimensions.}
    \label{fig:attention_architectures}
\end{figure} 

We utilize the multi-head attention (MHA) layer \cite{vaswani2017attention}. Each one of the $H$ heads performs attention over some low-dimensional embeddings derived from the tensors $Q$ and $V$, with the output embedding depth reduced to $D / H$. These independent attention heads have the capability to focus on different semantics of the input tensors. After performing attention across the specified axes $\mathcal{A}$, the final output is given by aggregating across the head outputs $o^{(h)}$:
\begin{equation}
    \begin{aligned}
o^{(h)} = \operatorname{Att}_{\mathcal{A}}(f_Q^{(h)}(Q), f_V^{(h)}(V), f_V^{(h)}(V)),
\\
\operatorname{MHA}_{\mathcal{A}}(Q, V) = f(\operatorname{Concat}(o_1, \dots, o_H)),
    \end{aligned}
    \label{eq:mha}
\end{equation}
where 
$f$ 
denotes a dense layer 
$\R^{X_Q \times T_Q \times D}
\rightarrow 
\R^{X_Q \times T_Q \times D}$ 
and the dense layers 
$f_Q^{(h)}$
and 
$f_V^{(h)}$
are linear maps
$\R^{X_Q \times T_Q \times D} 
\rightarrow  
\R^{X_Q \times T_Q \times D / H}$
and 
$\R^{X_V \times T_V \times D} 
\rightarrow  
\R^{X_V \times T_V \times D / H}$, respectively for each one of the $h \in {1, \dots, H}$ heads, where for our purposes we always assume that the keys and values tensors are the same size.  
Using these definitions, we now formulate our proposed attention methods.

\subsubsection{Self-attention (SA)}
\label{model:attention:sa}
First, we concatenate the audio tensor $A_\mathrm{SA}^{(0)}=Z_\mathrm{A}$ and the video $V_\mathrm{SA}^{(0)}=Z_\mathrm{V}$ tensors across the first axis
to form the $(M+G) \times T \times D$ tensor $Z_\mathrm{SA}^{(0)}$, the input to the first self-attention layer.

\textbf{Joint SA}:  Attention is performed jointly across space, time, and sources (see Figure \ref{fig:sa_block}). We express the $l$-th layer of a joint self-attention module as follows:
\begin{equation}
    \begin{aligned}
b^{(l)} =  &  \operatorname{MHA}_{\{\mathrm{M+G},\mathrm{T}\}}(Z_\mathrm{SA}^{(l-1)}, Z_\mathrm{SA}^{(l-1)}) + Z_\mathrm{SA}^{(l-1)},  \\
Z_\mathrm{SA}^{(l)} = & \operatorname{LN}( f^{(l)} ( \operatorname{Dropout}(b^{(l)}) ) + b^{(l)} ),
\end{aligned}
    \label{eq:joint_sa}
\end{equation}
where $f^{(l)}$ is a dense layer, $\operatorname{LN}$ is layer normalization \cite{ba2016layernorm} and $\operatorname{Dropout}$ denotes a dropout layer \cite{srivastava2014dropout}. We define the sequence of operations in (\ref{eq:joint_sa}) as $Z_\mathrm{SA}^{(l)} = \operatorname{SA}_{\{\mathrm{M+G},\mathrm{T}\}}(Z_\mathrm{SA}^{(l-1)})$ where the self-attention is performed across the joint sources-and-spatial dimension $\mathrm{M+G}$ and the time axis $\mathrm{T}$. The final representation $z$, after the repetition of $L$ self-attention blocks, for all $M$ sources, is obtained by slicing and performing attentional pooling \cite{tzinis2021into} across the time axis of $\widehat{z}$:
\begin{equation}
    \begin{aligned}
    z = & \operatorname{MHA}_{\{\mathrm{T}\}}
    (
        {\textstyle \sum}_t^{T}\widehat{z}_t, \widehat{z}
    )
    \in \R^{M \times D}, \enskip 
     \widehat{z} =  A_\mathrm{SA}^{(L)}=Z_\mathrm{SA}^{(L)}[1:M] \in \R^{M \times T \times D}.
    \end{aligned}
    \label{eq:final_jointselfattetnion}
\end{equation}

\textbf{Separable SA}: 
The $l$-th separable self-attention block can be expressed using the SA module defined in (\ref{eq:joint_sa}):
\begin{equation}
    \begin{gathered}
a^{(l)} =  \operatorname{SA}_{\{\mathrm{T}\}}(Z_\mathrm{SA}^{(l-1)}[1:M]), \enskip
v^{(l)} =  \operatorname{SA}_{\{\mathrm{T}\}}(Z_\mathrm{SA}^{(l-1)}[M:M+G]), \\
Z_\mathrm{SA}^{(l)} =  \operatorname{SA}_{\{\mathrm{M+G}\}}\left( \operatorname{Concat}(a^{(l)}, v^{(l)}) \right).
    \end{gathered}
    \label{eq:separable_sa}
\end{equation}
The final audio-visual representation is obtained through attentional pooling and slicing as before (see also Figure \ref{fig:separable_cma_block}).

\subsubsection{Cross-modal attention (CMA)}
\label{model:attention:cma}
In this attention layer we keep the audio and the video modality tensors separate, and we perform queries from one modality to another. Formally, the input to the stacked CMA blocks is a pair of an audio $A^{(0)}_\mathrm{CMA} \in \R^{M \times T \times D}$ and a video $V^{(0)}_\mathrm{CMA} \in \R^{G \times T \times D}$ feature tensors.

\textbf{Joint CMA}: We perform a directional attention from the audio (video) modality tensor to the video (audio) tensor, attending across both sources and time (space and time) axes. Formally, at the $l$-th layer we have the following sequence of operations for the directional attention where we use as a query the audio modality, also illustrated in Figure \ref{fig:cma_block}:
\begin{equation}
    \begin{gathered}
a_1^{(l)} =  \operatorname{MHA}_{\{\mathrm{G}, \mathrm{T}\}}(A_\mathrm{CMA}^{(l-1)}, V_\mathrm{CMA}^{(l-1)}), \enskip
a_2^{(l)} =  \operatorname{LN}(a_1^{(l)} + A_\mathrm{CMA}^{(l-1)}), \\
A_\mathrm{CMA}^{(l)} =  \operatorname{LN} (f ( \operatorname{Dropout}(a_2^{(l)}) ) + A_\mathrm{CMA}^{(l-1)} ). \\
    \end{gathered}
    \label{eq:joint_cma_audio}
\end{equation}
For the other direction, we modulate the video features $v^{(l)}$ using the audio features $a^{(l)}$ by swapping $A_\mathrm{CMA}$ and $V_\mathrm{CMA}$ and attending over $\{\mathrm{M},\mathrm{T}\}$ in (\ref{eq:joint_cma_audio}).

We define (\ref{eq:joint_cma_audio}) as
$
    A_\mathrm{CMA}^{(l)}, V_\mathrm{CMA}^{(l)}
    =
    \noindent
    \operatorname{CMA}_{\{\mathrm{M|G}, \mathrm{T}\}}
    (A_\mathrm{CMA}^{(l-1)}, V_\mathrm{CMA}^{(l-1)}),
$
where each cross-modal attention is performed across the dimension of audio sources $M$ or spatial locations $G$ (denoted in our notation as "$\mathrm{M|G}$") and the time axis $\mathrm{T}$. The output audio-visual embedding $z$ contains information for all $M$ sources and is obtained after the repetition of $L$ cross-modal attention blocks via attentional pooling across time (\ref{eq:final_jointselfattetnion}) on the output audio tensor 
$\widehat{z} = A_\mathrm{CMA}^{(l)} \in \R^{M \times T \times D}$.

\textbf{Separable CMA}: Similar to Section \ref{model:attention:sa}, we can reduce the space complexity of the proposed CMA layer by first performing self-attention across the time axis for each modality separately, then performing CMA across the remaining axis (i.e. sources or spatial locations) as shown next, also illustrated in Figure \ref{fig:separable_cma_block}:
\begin{equation}
    \begin{gathered}
a^{(l)} =  \operatorname{SA}_{\{\mathrm{T}\}}(A_\mathrm{CMA}^{(l-1)}), \enskip
v^{(l)} =  \operatorname{SA}_{\{\mathrm{T}\}}(V_\mathrm{CMA}^{(l-1)}), \\
A_\mathrm{CMA}^{(l)}, V_\mathrm{CMA}^{(l)} = \operatorname{CMA}_{\{\mathrm{M|G}\}}(a^{(l)}, v^{(l)}).
    \end{gathered}
    \label{eq:separable_cma}
\end{equation}

\subsection{Audio-visual on-screen sound classifier}
\label{model:classifier}
For each estimated source $\hat{s}_m$, AudioScopeV2 predicts the probability $\hat{y}_m$ that it originates from an on-screen object.
These probabilities are computed using the extracted audio-visual representation from the output $z$ of our attention-based models
for the self-attention and cross-modal attention encoders. Specifically, for each source $m$, we feed the audio-visual embedding $z_m \in \R^{D}$ through a dense layer $f_z$ tied across sources to produce logits $\hat{\ell}_m$, and then apply an element-wise sigmoid activation $\sigma$ to compute the audio-visual coincidence probability $\hat{y}_m$.
The final on-screen waveform estimate $\hat{x}^\mathrm{on}$ is produced using these probabilities as soft weights and multiplied with the corresponding estimated sources:
\begin{equation}
\begin{gathered}
    \hat{\ell}_m = f_z (z_m),
    \enskip
    \hat{y}_m = \sigma \Big( \hat{\ell}_m \Big) \in [0, 1],
    \enskip 
    \label{eq:onscreen_estimate}
    \hat{x}^\mathrm{on} =
    {\textstyle \sum}_{m=1}^{M}
    \hat{y}_m \hat{s}_m.
\end{gathered}
\end{equation}

\subsection{Training procedure}
\label{ssec:training_procedure}

To train the separation model, we use MixIT \cite{MixITNeurIPS}. Given two reference mixtures $r_1,r_2\in\mathbb{R}^{T'}$, $M$ separated sources $\hat{s}$ predicted from the MoM $x=r_1+r_2$, and a signal-level training loss $\mathcal{L}$, MixIT infers an optimal $2\times M$ binary mixing matrix $A$ where each column sums to one. This mixing matrix assigns each estimated source $s_m$ to one of the reference mixtures $r_1$ or $r_2$:
\begin{align}
\mathcal{L}_\mathrm{MixIT}
\left(
    {r}, \hat{s}
\right)
=
\min_{{A}\in\mathbb{B}^{2 \times M}} \, & {\textstyle \sum}_{n=1}^2\mathcal{L}
    \left(
        r_n, [{A} \hat{s}]_n
    \right).
\label{eq:mixit}
\end{align}
For $\mathcal{L}$, we use the negative thresholded SNR loss \cite{MixITNeurIPS, tzinis2021into}.

We use purely unsupervised training, which uses batches composed of \emph{noisy-labeled on-screen} (NOn) examples. Each NOn example consists of the video frames and audio for a primary input 5-second video clip,
where additional audio from another random 5-second video clip 
serves as synthetic off-screen audio that is mixed with the primary video soundtrack. These examples provide noisy pseudo-labels, because even after optimal MixIT combination, the primary soundtrack may contain off-screen background noise.

For training AudioScopeV2's audio-visual on-screen classifier with NOn examples, we use the \emph{active combinations} loss $\mathcal{L}_{\operatorname{AC}}$ \cite{tzinis2021into} computed between the pseudo-label assignments $y$ provided by MixIT and the predictions $\hat{y}$ of the classifier. $\mathcal{L}_{\operatorname{AC}}$ corresponds to the minimum cross-entropy loss $\mathcal{L}_\mathrm{CE}$ over all settings $\powerset_{\geq 1} \left( \mathbb{B}^M \right)$ of the labels such that at least one label is $1$ (equivalent to at least one source appearing on-screen):
\begin{equation}
    \label{losses:active_combinations}
    \mathcal{L}_{\operatorname{AC}}\left( y, \hat{y} \right) = 
    \min_{ \ell \in \powerset_{\geq 1} \left( \mathbb{B}^M \right) }
    {\textstyle \sum}_m
    \mathcal{L}_\mathrm{CE}(\ell_m, \hat{y}_m).
\end{equation}

\section{Experimental Framework}

\subsection{Data preparation and labeling}
\label{ssec:data_prep}

For our open-domain experiments we use 
Flickr Creative Commons 100 Million Dataset (YFCC100M) \cite{thomee2016yfcc100m} CC-BY videos respecting published train/validation/test splits \cite{yfcc100m_github}. Instead of using only a filtered subset of the data, as previous state-of-the-art methods suggested \cite{tzinis2021into}, our proposed training recipe is able to leverage unrestricted open-domain datasets 
by
using better pre-trained separation models.

The unfiltered training data consists of $\approx 1600$ hours, and we extract 5-second clips with a hop of 1 second (4.85M total clips). We also gathered human annotations for 5-second clips from unfiltered videos from the train, validation, and test splits. 
The count of total clips rated, unanimously-rated on-screen-only clips, and unanimously rated off-screen clips were 20000/480/4664 for training, 6500/109/1421 for validation, and 3500/43/762 for test. 
Notice that during training we dynamically create mixtures of pairs of these clips, so the effective number of unique examples is
$\mathcal{O}(10^{13})$. We also experiment with faster video frame rate of 16 frames per second (FPS), instead of 1 FPS used by
AudioScope
\cite{tzinis2021into}.

\subsection{Training details}
\label{ssec:training_details}
Both audio and visual embedding networks were pre-trained on AudioSet \cite{AudioSet} for unsupervised coincidence prediction \cite{jansen2020coincidence}. We also found that freezing these networks during training leads to better results. Also, instead of training the separation model from scratch, we pre-train the separation model with MixIT on unfiltered audio-only MoMs drawn from YFCC100M for 3.6M steps, which also significantly boosted the performance of our models. We use $L=4$ stacked proposed layers of joint/separable SA/CMA (ablation studies can be found in the supplementary material). 
All models were trained on 32 Google Cloud TPU v3 cores with the Adam optimizer \cite{kingma2014adam}, batch size $128$, and learning rate $10^{-4}$.

\subsection{Evaluation datasets}

In order to compare with the current state-of-the-art, we use the AudioScope dataset splits \cite{tzinis2021into} provided online \cite{audioscope_yfcc100m_github}, which were drawn from a subset of YFCC100M filtered by an unsupervised coincidence model \cite{jansen2020coincidence}. These datasets (we refer to them as \textit{filtered off-screen background}) contain two kinds of examples: \textit{on-screen MoMs}, where additional off-screen audio was injected into the soundtrack of a unanimously-rated on-screen-only video (i.e.\ input audio $x=x^\mathrm{on}+x^\mathrm{off}$), and \textit{off-screen MoMs}, where additional audio was mixed into the soundtrack of a unanimously-rated off-screen-only video (input audio $x=x^\mathrm{off}$). 

Similarly, we construct new evaluation sets of MoMs using videos from unfiltered validation and test splits of YFCC100M, which we call
\textit{unfiltered random background}. Instead of using off-screen-only audio, our dataset uses randomly sampled audio from all unfiltered videos as synthetic off-screen background for unanimously-rated on-screen-only or off-screen-only videos
(see Section \ref{ssec:data_prep}).

\subsection{Evaluation metrics and calibration}
\label{ssec:metrics_and_calibration}
For on-screen examples with input audio $x=x^\mathrm{on}+x^\mathrm{off}$, we report the reconstruction fidelity of on-screen estimates $\hat{x}^\mathrm{on}$ (\ref{eq:onscreen_estimate}) using \emph{signal-to-noise ratio} (SNR) in dB.  For off-screen examples with input audio $x=x^\mathrm{off}$ where we know that no audio originates from on-screen objects, we measure the ability of our models to suppress off-screen sources using \emph{off-screen suppression ratio} (OSR).
\begin{equation}
    \mathrm{SNR}({x}^\mathrm{on}, \hat{x}^\mathrm{on}) 
    = 20\log_{10} \frac{\|x^\mathrm{on}\|}{ \|x^\mathrm{on} - \hat{x}^\mathrm{on}\|}, 
    \mathrm{OSR}(x, \hat{x}^\mathrm{on})
    =
    20\log_{10}
    \frac
        {\|x\|}
        {\|\hat{x}^\mathrm{on}\|}.
    \label{eq:snrosr}
\end{equation}
OSR measures the power reduction of the on-screen estimate $\hat{x}^\mathrm{on}$ relative to the input audio, and is only measured on examples where the input audio $x=x^\mathrm{off}$ is entirely off-screen.
Prior work \cite{tzinis2021into} has also used scale-invariant signal-to-noise ratio (SI-SNR) \cite{LeRoux2018a} instead of SNR; however, 
this allows a model to obtain large OSR, without sacrificing SI-SNR, by scaling down its estimates.  

\begin{figure}[h]
  \centering
  \includegraphics[width=0.6\linewidth]{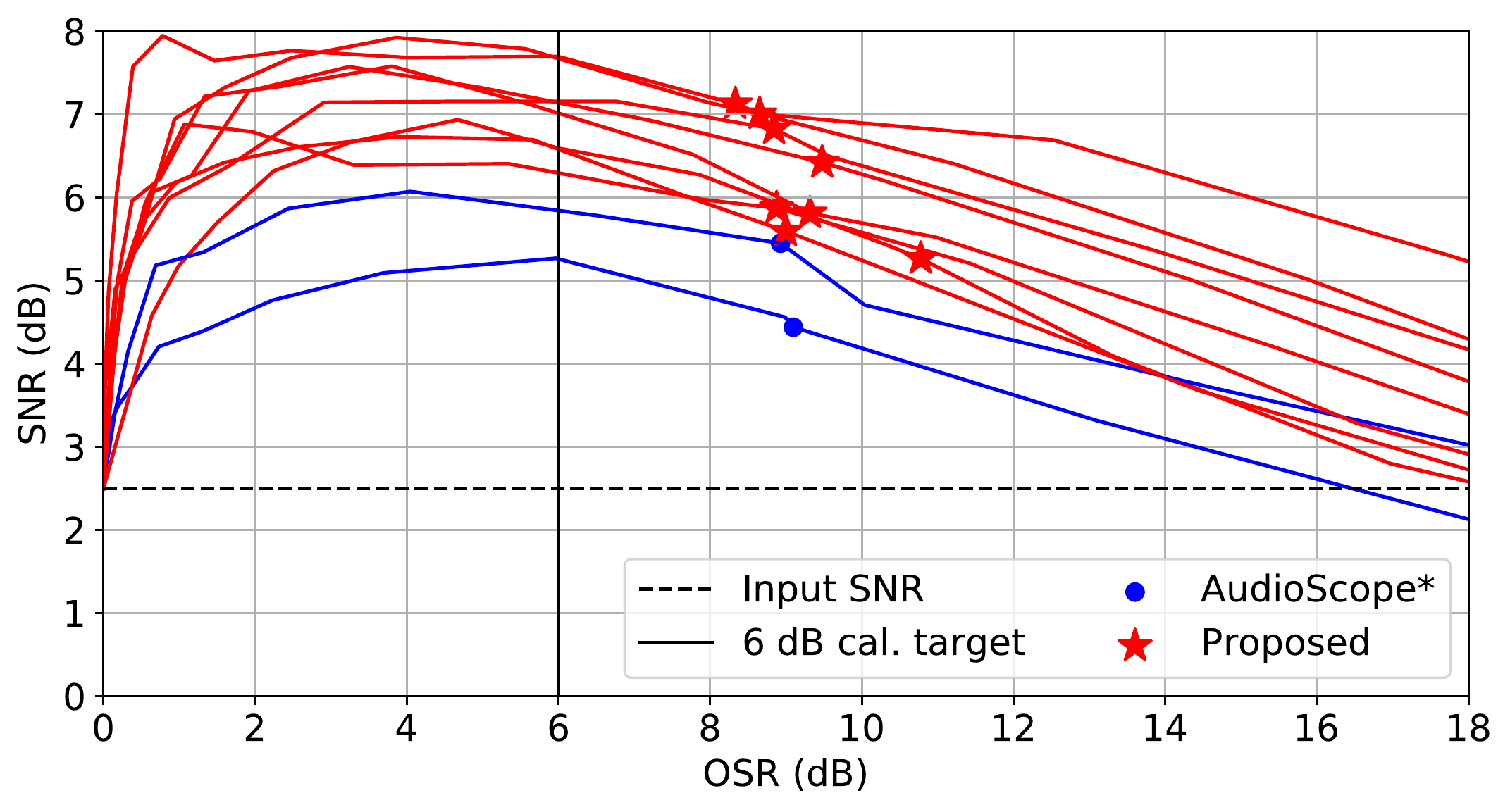}
\caption{OSR versus SNR curves when calibration offset $\theta$ in (\ref{eq:onscreen_estimate_calibrated}) is varied, on unfiltered random background test set for all models in Table \ref{tab:results} (except for AudioScope \cite{tzinis2021into}, which achieves 23.4 dB OSR / 1.1 dB SNR for 1 FPS).
} 
  \label{fig:osr_vs_snr}
\end{figure}

Both OSR and SNR are important, but there is an inherent trade-off between them, since OSR can always be increased by scaling down the output, at the expense of SNR.  This makes it difficult to compare models that have different operating points in this trade-off, a problem not addressed by \cite{tzinis2021into}. %
To illustrate this, Figure \ref{fig:osr_vs_snr} plots OSR versus SNR for the models we consider in this paper. Notice that each model achieves a different operating point.
Because of these differences, SNR cannot be meaningfully compared across models without considering OSR. For example, one of our proposed models achieves a lower SNR of about 5.3 dB at 10.8 dB OSR, compared to an AudioScope* model that achieves a higher SNR of 5.5 dB at 8.9 dB OSR. It is difficult to say which model is better.

To solve this problem, we propose a novel calibration method for these on-screen separation models, where we adjust a bias in the classifiers to achieve a target average OSR. This is akin to 
choosing a threshold for a detector to achieve a target false positive rate.  
For our procedure, we define a calibrated on-screen estimate given by adding a global scalar offset $\theta$ to the on-screen logits $\hat{\ell}_{1:M}$ (\ref{eq:onscreen_estimate}):
\begin{equation}
    \tilde{x}^\mathrm{on}(\theta)
    =
    {\textstyle \sum}_{m} \hat{s}_{m} \sigma(\hat{\ell}_{m} + \theta).
    \label{eq:onscreen_estimate_calibrated}
\end{equation}
The offset $\theta$ is tuned such that the median OSR (\ref{eq:snrosr}) across all $N_\mathrm{off}$ off-screen examples,
$
    \mathrm{med}_{j=1}^{N_\mathrm{off}}
    \mathrm{OSR}
    \left[
        x_j,
        \tilde{x}^\mathrm{on}_j(\theta)
    \right],
$
is equal to a desired $\mathrm{OSR}_\mathrm{target}$. The curves in Figure \ref{fig:osr_vs_snr} illustrate the effect on OSR and SNR when $\theta$ is varied, and are akin to receiver operating characteristic (ROC) or precision-recall curves.

In practice, $\theta$ has a monotonic relationship to OSR. As $\theta$ tends towards $0$ (inversely $\infty$), the on-screen probabilities $\hat{y}_m$ tend towards $0$ (inversely $1$), and thus OSR approaches $\infty$ (inversely $0$) dB. Because of this property, optimization of $\theta$ is very simple, and can be accomplished efficiently via binary search.

For our results, we use $\mathrm{OSR}_\mathrm{target}=6$dB, which corresponds to all off-screen sources sounding as if they are twice as far away.
To choose the early stopping point, we evaluate all models on the \emph{unfiltered random background} validation data. Running calibration on each step of model training would be prohibitively expensive. Thus, we choose the point which maximizes the minimum of SNR and OSR, favoring models which are not strongly biased towards any metric.

We also report the performance of 
classification
using weighted area under the curve of the ROC (AUC-ROC), where the weight for each probability $\hat{y}_m$ is the normalized power of the corresponding source $\hat{s}_m$: $\|\hat{s}_m\|_2^2 / \left({\textstyle \sum}_{m'} \|\hat{s}_{m'}\|_2^2\right)$.
\section{Results}
\label{results}
\subsection{Open-domain on-screen separation}

\begin{table}[h]
    \centering
    \caption{Evaluation results
    for filtered off-screen background (from \cite{tzinis2021into}) and unfiltered random background (our new proposed) test sets at 1 and 16 FPS. Separate models are trained for 1 FPS and 16 FPS.  For each model, calibration to 6 dB OSR is performed separately on the filtered and unfiltered test sets. ``PT'' indicates separation model pre-training, ``Filt.''  means training on filtered data, and ``Complexity'' is the theoretical complexity of each AV-alignment module.
    }
    \label{tab:results}
\scalebox{0.81}{
\begin{tabular}{
l
@{\hskip 20\tabcolsep}
c
@{\hskip 20\tabcolsep}
l
@{\hskip 1\tabcolsep}
c
c
@{\hskip 5\tabcolsep}
c
c
@{\hskip 5\tabcolsep}
c
c
@{\hskip 5\tabcolsep}
c
c
@{\hskip 5\tabcolsep}
c
c
}
\multicolumn{2}{c}{} & &
&
&\multicolumn{4}{c}{Filtered \cite{tzinis2021into}}
&
\multicolumn{4}{c}{Unfiltered (new proposed)}\\
\cmidrule(r{\dimexpr 5\tabcolsep}){6-9} 
\cmidrule(r{\dimexpr 1\tabcolsep}){10-13}
\multicolumn{2}{c}{} & &
&
&\multicolumn{2}{c}{1 FPS}
&
\multicolumn{2}{c}{16 FPS}&
\multicolumn{2}{c}{1 FPS}
&
\multicolumn{2}{c}{16 FPS}
\\
\cmidrule(r{\dimexpr 5\tabcolsep}){6-7} 
\cmidrule(r{\dimexpr 5\tabcolsep}){8-9}
\cmidrule(r{\dimexpr 5\tabcolsep}){10-11} 
\cmidrule(r{\dimexpr 1\tabcolsep}){12-13}
\multicolumn{2}{l}{AV alignment}
&Complexity 
&PT & Filt.
&{\bf SNR}
&{\bf AUC}
&{\bf SNR}
&{\bf AUC}
&{\bf SNR}
&{\bf AUC}
&{\bf SNR}
&{\bf AUC}
\\
\midrule
\multicolumn{5}{l}{\multirow{1}{*}{No processing ($\hat{x}^\mathrm{on}=x$ with 0dB OSR)}}  &  4.4  & -- & 4.4 & -- & 2.5  & -- & 2.5 & -- \\
\multicolumn{5}{l}{\multirow{1}{*}{No processing ($\hat{x}^\mathrm{on}=x / 2$ with 6dB OSR)}}  &  4.7  & -- & 4.7 & -- & 4.1  & -- & 4.1 & -- \\
\midrule
\multicolumn{2}{l}{\multirow{1}{*}{AudioScope \cite{tzinis2021into}}} & $\mathcal{O}(T  M  G)$ &  & \checkmark  & 6.0 & 0.79 & -- & -- & 2.7  & 0.69 &-- &--  \\
\multicolumn{2}{l}{\multirow{1}{*}{AudioScope*}} & $\mathcal{O}(T  M  G)$ & \checkmark &  & 8.2 & 0.80 & 5.9 & 0.77 & 5.8  & 0.78 & 5.2 & 0.71 \\ \midrule
\multicolumn{1}{l}{\multirow{2}{*}{SA}} & \multirow{1}{*}{Joint} & \multirow{1}{*}{$\mathcal{O}(T^2  [M + G]^2)$} & \checkmark & & \textbf{10.0} & 0.84 & 9.9 & \textbf{0.86} & 7.2 & 0.82 & {\bf 7.7} & 0.83 \\
 & \multirow{1}{*}{Sep.} & \multirow{1}{*}{$\mathcal{O}(T^2 + [M + G]^2)$} & \checkmark &  & 9.6 & 0.84 & 8.2 & 0.83 & 6.6  & 0.78 & 6.6 & 0.80 \\
\multicolumn{1}{l}{\multirow{2}{*}{CMA}} & \multirow{1}{*}{Joint} & $\mathcal{O}(T^2 M  G)$ & \checkmark &  & \textbf{10.0} & \textbf{0.88} & \textbf{10.0} & 0.85 & {\bf 7.3} & {\bf 0.83} & {\bf 7.7} & {\bf 0.84} \\
 & \multirow{1}{*}{Sep.} & $\mathcal{O}(T^2 + M G)$ & \checkmark &  & 9.5 & 0.83 & 9.3 & 0.82 & 6.4 & 0.78 & 7.1 & 0.80 \\
\midrule
\end{tabular}}
\end{table}

Results are shown in 
Table \ref{tab:results}
for the filtered off-screen background dataset from \cite{tzinis2021into}, and our new and challenging
unfiltered random background dataset.
We include two ``no processing'' baselines, with the on-screen estimate equal to the input audio $\hat{x}^\mathrm{on}=x$ (0 dB OSR), or half the input $\hat{x}^\mathrm{on}=x/2$ (6 dB OSR).

On the filtered test set, our proposed models 
significantly outperform the previous state-of-the-art AudioScope model \cite{tzinis2021into} trained on filtered data, by more than $4$dB in terms of SNR and 0.09 in AUC-ROC.
Training on unfiltered data, including pre-training the separation model on this data (AudioScope*), improves over the baseline AudioScope model trained on filtered data (6.0 dB $\rightarrow$ 8.2 dB).

On our newly introduced unfiltered random background dataset, AudioScope \cite{tzinis2021into} trained on mismatched filtered data suffers from poor generalization, achieving only 2.7 dB SNR, which is worse than doing no processing 
(4.1 dB SNR). This demonstrates the limitation of training on filtered data. For our proposed models, joint CMA yields the best improvements over AudioScope* trained with matched unfiltered data (5.8 dB $\rightarrow$ 7.7 dB SNR and 0.78 $\rightarrow$ 0.84 AUC-ROC). Also, notice that the much more efficient separable versions of SA and CMA only suffer minor degradation compared to the joint versions (7.2 dB $\rightarrow$ 6.6 dB for SA, 7.3 dB $\rightarrow$ 6.4 dB for CMA).
Our proposed models can perceive and leverage higher-frequency dynamics from the audio-visual scene and thus, using a higher frame rate of 16 FPS provides improvement of up to 0.7 dB SNR. 
On the other hand, AudioScope* seemingly cannot scale to higher frame-rates (8.2 dB $\rightarrow$ 5.9 dB for filtered and 5.8 dB $\rightarrow$ 5.2 dB for unfiltered, going from 1 FPS to 16 FPS), presumably because its audio-visual alignment is limited by only using shallow spatio-temporal attention. 
 It is also possible that filtering with the audio-visual coincidence model \cite{jansen2020coincidence}, which averaged coincidence scores over frames at 1 FPS, may have biased towards video clips with always-visible sounding objects.

\subsection{Computational efficiency
}
\label{sec:results:efficiency}
\begin{figure}[htb!]
  \centering
  \includegraphics[width=0.65\linewidth]{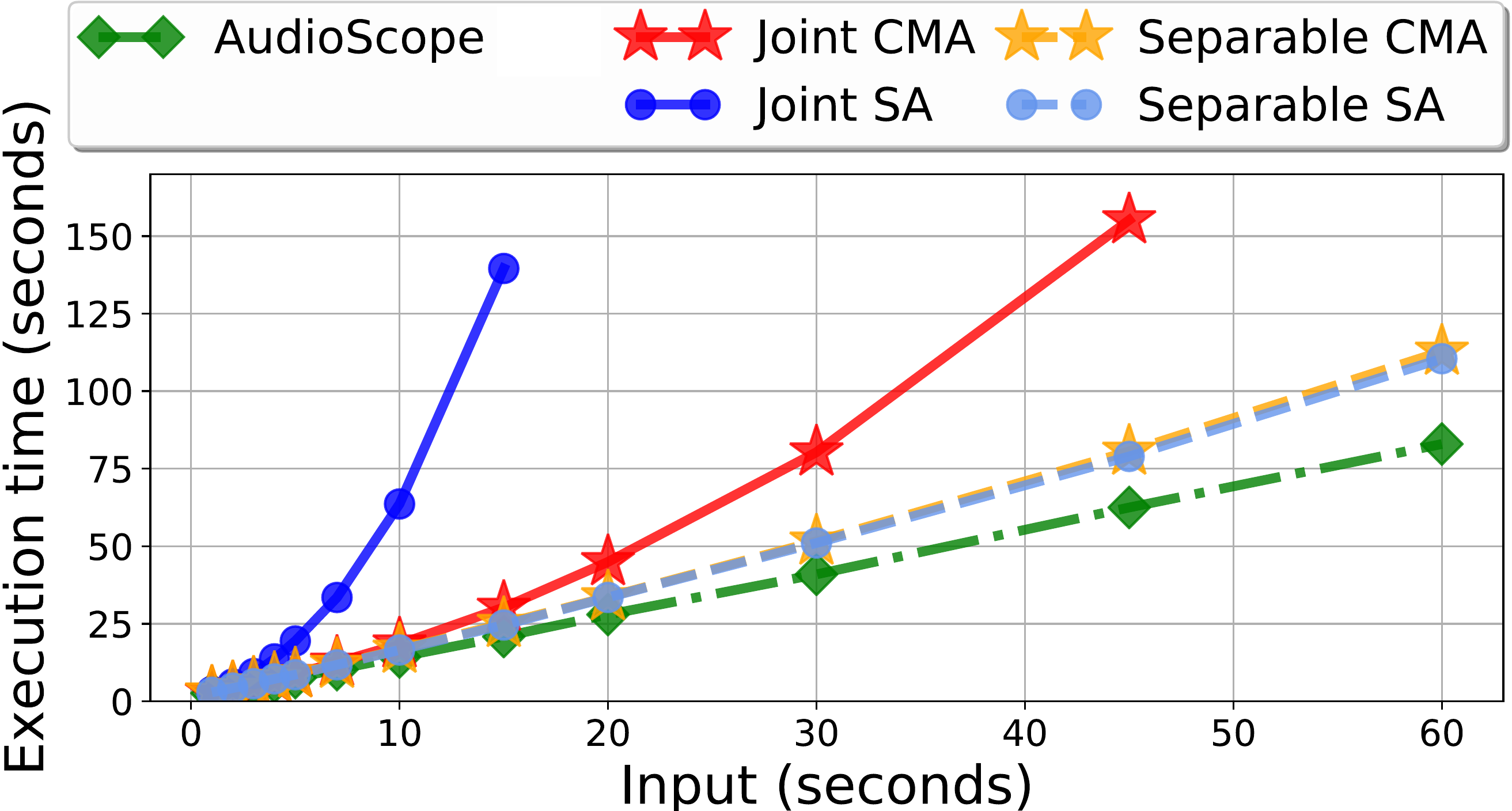}
\caption{Computation requirements AudioScope 
\cite{tzinis2021into} and our proposed models
with separable and joint SA and CMA for different video input lengths at 16 FPS. All measurements were taken on a 
machine with 16 GB of RAM and 2 
Intel Xeon CPU @ 2.20GHz cores. Input duration is increased until out-of-memory error.} 
  \label{fig:efficiency_main}
\end{figure}

In theory, joint SA has complexity $\mathcal{O}(T^2  [M + G]^2)$, which scales poorly relative its separable version with complexity $\mathcal{O}(T^2 + [M + G]^2)$. Joint CMA has a slightly better complexity $\mathcal{O}(T^2 M G)$ but still lags behind its separable counterpart $\mathcal{O}(T^2 + M G)$. However, it is important to evaluate how the computation scales in practice. The execution time versus input video length for all models is depicted in Figure \ref{fig:efficiency_main}. 
The efficiency of the proposed separable self- and cross-modal attention becomes apparent for longer videos. 
These separable variants perform
comparably to the much more complex joint CMA and joint SA models (see Table \ref{tab:results}) but have computational requirements on par with the simpler AudioScope \cite{tzinis2021into}.

\section{Conclusion}
\label{conclusion}
We identified several issues with 
the previous state-of-the-art open-domain audio-visual on-screen sound separation model, AudioScope \cite{tzinis2021into}.
These issues include oversimplicity of shallow attention used for audio-visual alignment, lack of generalization due to filtering video data with an unsupervised audio-visual coincidence model \cite{jansen2020coincidence}, and inability to specify the trade off between on-screen reconstruction and off-screen suppression. We proposed more sophisticated self- and cross-modal attention architectures, along with more efficient separable versions, that improve performance and are able to leverage higher video frame rates.
To address lack of generalization, we provide annotations for a new dataset constructed from YFCC100M \cite{thomee2016yfcc100m} that is unfiltered by the audio-visual coincidence model. As a result, this dataset is more diverse and representative of in-the-wild video data. Finally, we proposed a calibration procedure that allows any on-screen separation model to be tuned for a specific level of off-screen suppression, which allows more control over model behavior and much easier comparison between different models.
Using our calibration procedure, our results show that our proposed architecture is able to generalize to our more challenging test set and achieve clear improvements over the previous state-of-the-art AudioScope model \cite{tzinis2021into}.

\bibliographystyle{splncs04}
\bibliography{egbib}

\title{AudioScopeV2: Audio-Visual Attention Architectures for Calibrated Open-Domain On-Screen Sound Separation\\ (Supplementary Material)} %

\titlerunning{AudioScopeV2: Audio-Visual Attention for Calibrated On-Screen Separation}
\author{Efthymios Tzinis\inst{1,2}\thanks{Work done during an internship at Google Research.}\and
Scott Wisdom\inst{1} \and
Tal Remez\inst{1}\and
John R.\ Hershey\inst{1}}
\authorrunning{E. Tzinis et al.}
\institute{
Google Research
\and
University of Illinois Urbana-Champaign
\\
\email{etzinis2@illinois.edu}, \email{\{scottwisdom,johnhershey\}@google.com}}
\maketitle

\section*{Outline of the supplementary material}

\begin{itemize}
    \item Section \ref{sec:demos}: video demos
    \item Section \ref{sec:calibration}: analysis of calibration
    \item Section \ref{sec:ablations}: ablations
    \item Section \ref{sec:eval}: evaluation on restricted-domain datasets
    \item Section \ref{sec:visual}: visualizations of attention maps
\end{itemize}

\section{Video demos}
\label{sec:demos}

Please see our project website\footnote{\url{google-research.github.io/sound-separation/papers/audioscope-v2}} for video demos using our proposed AudioScopeV2 and AudioScope \cite{tzinis2021into}. For these demos, we run the following models trained unsupervised on our proposed unfiltered YFCC100M \cite{thomee2016yfcc100m} data: joint and separable SA at 16 FPS, joint and separable CMA at 16 FPS, and AudioScope* (improved version of AudioScope, as described in the main paper) at 16 and 1 FPS.
We also use an AudioScope model \cite{tzinis2021into} at 1 FPS, which was trained on filtered YFCC100M.
We provide demos on two types of examples: synthetic mixtures of mixtures (MoMs) from the unfiltered random background test set, and single real videos drawn from the test split of unfiltered YFCC100M.

The demos highlight a variety of interesting cases.  
\begin{itemize}
\item Real Example 1 shows a close-up of a child talking on-screen with strong non-stationary noise in the background; the separable SA models do a remarkable job of suppressing the background noise.
    \item In real Example 2, there is a child and an adult presumably talking on screen, in the midst of loud off-screen speech from a news broadcast; the separable SA models successfully suppress the off-screen voice while preserving the presumed on-screen subjects of the video, despite their faces being obscured, whereas the AudioScope models yield inconsistent results. Taken together these examples illustrate that the models can selectively preserve or block speech depending on whether the talkers are on-screen, even without a frontal view of the on-screen faces; presumably the model is able to use context and other cues to infer which voices correspond to what appears on-screen.
    \item In real Example 3, a child is playing in a pile of dried leaves with ambient noise in the background.
    \item In real Example 4, a rocking horse ridden by a child is impacting the walls in a hallway, with some clicking and ambient background noise. In both of these examples, the models are able to differentially suppress the background noise and preserve the on-screen noises, even though both are noise-like sounds.
    \item The other real examples show operation in low-light conditions (Example 5), selective enhancement of a baseball hit while suppressing background voices (Example 6), and selective enhancement of non-speech eating sounds (Example 7).
\end{itemize}

In the synthetic demos, Examples 2 and 6 show the model selectively removing or preserving stationary noise depending on the visual input. 
Synthetic Example 2 shows a person talking on-screen, with off-screen mechanical noise; the models substantially reduce the mechanical noise. 
Synthetic Example 6 has on-screen rain with off-screen wind noise and voices. 

Overall the models show a variety of interesting emergent behaviors, such as preserving inferred on-screen sources, and future work will require a more systematic analysis of conditions amenable to good performance.

\section{Analysis of calibration}
\label{sec:calibration}

In this section, we analyze the effect of calibration on the YFCC100M unfiltered test data. We were motivated by the ``no processing'' baseline providing some counter-intuitive results. As a reminder, we reported ``no processing'' baselines in Table 1 of our main paper at 0 dB and 6 dB OSR. The 0 dB OSR model just outputs the input audio $x$ as the on-screen estimate $\hat{x}^\mathrm{on}$. To calibrate the ``no processing'' model to 6 dB OSR, we simply use one half the input audio $x$ as the on-screen estimate $\hat{x}^\mathrm{on}$.

\begin{figure}[htb!]
  \centering
  \includegraphics[width=0.49\linewidth]{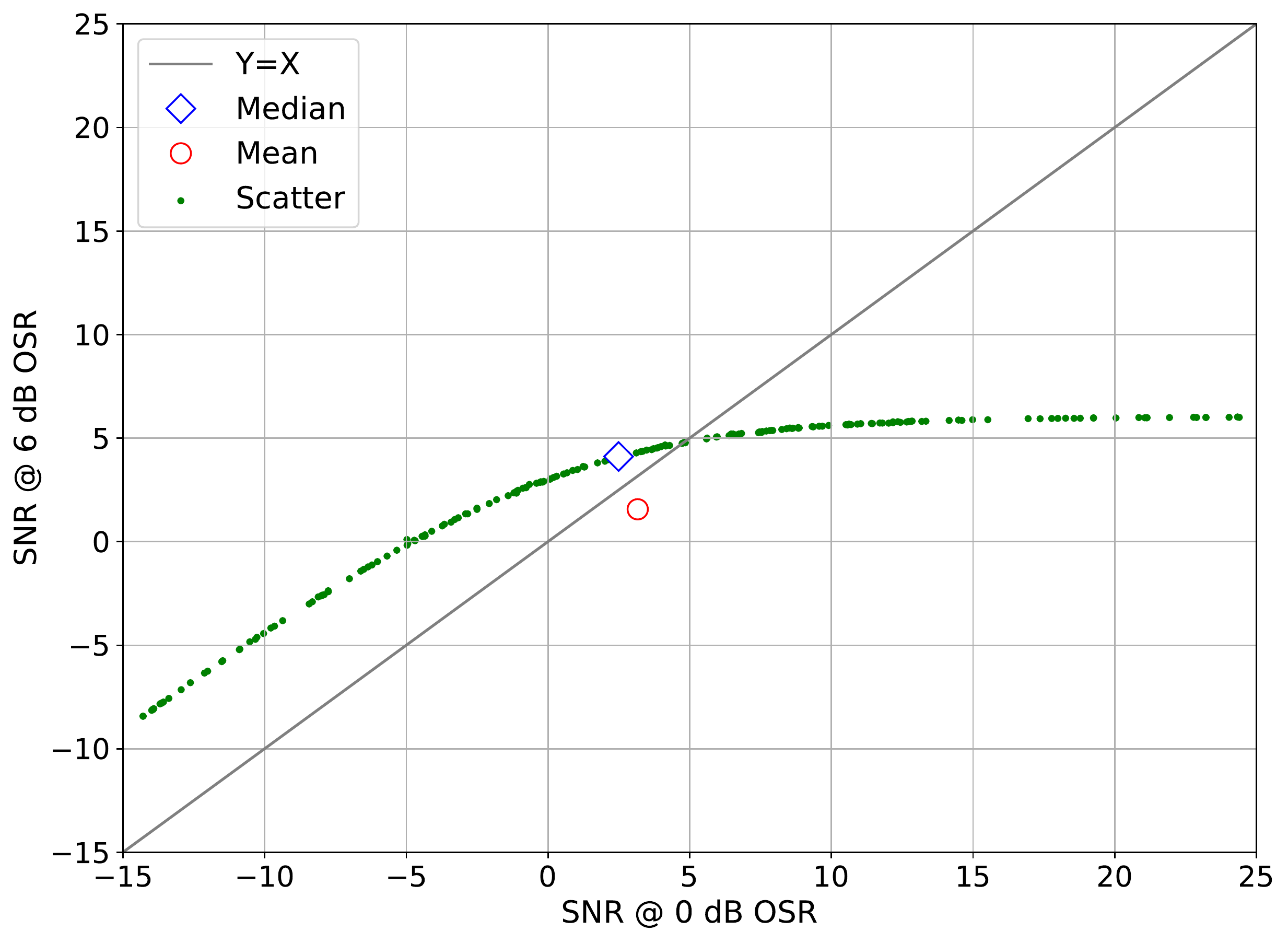}
  \includegraphics[width=0.49\linewidth]{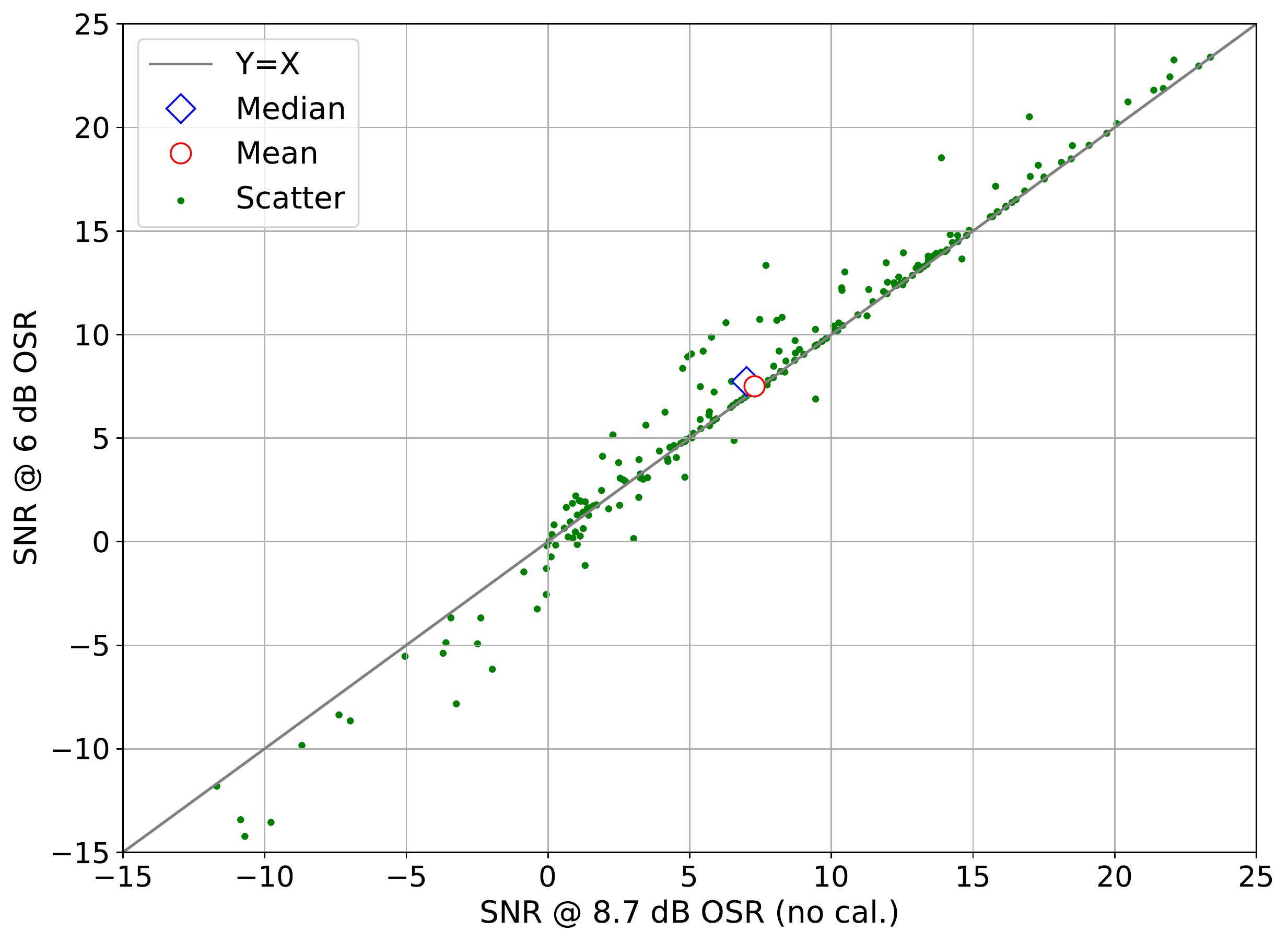}
  \caption{Left: scatter plot of ``no processing'' baseline SNRs at 0 dB OSR versus SNRs at 6 dB OSR. Right: scatter plot of the joint CMA model's SNRs before calibration (8.7 dB OSR) versus SNR at 6 dB OSR.
  }
  \label{fig:calibration_scatter}
\end{figure}

Counter-intuitively, the 0 dB OSR model achieves a lower median SNR than the 6 dB OSR model. To understand this effect, the left panel of Figure \ref{fig:calibration_scatter} shows a scatter plot of individual examples for 0 dB OSR versus 6 dB OSR. We can see that the median for 6 dB OSR is increased, at the expense of limiting the maximum attainable SNR to 6dB, and results in a drop in the mean SNR.

We also provide a similar plot for the joint CMA AudioScopeV2's configuration, in the right panel of Figure \ref{fig:calibration_scatter}, which plots SNR for the uncalibrated model, which achieves 8.7 dB OSR, versus SNR for the model calibrated at 6 dB OSR, following the procedure in Section 4.4 in the main paper. Note that in this case, in contrast to the ``no processing'' plot, the calibration is adjusting from higher OSR to lower OSR. Unlike the ``no processing'' case, the calibration procedure on AudioScopeV2 does not cause a drop in mean SNR, and improves SNR performance for examples that had greater than 0dB SNR in the original model.  SNR is decreased for examples that had less than 0dB SNR for the uncalibrated model, but the SNR-limited behavior of the calibrated ``no processing'' model is avoided.

\section{Ablations}
\label{sec:ablations}

This section presents several ablations that we could not include in the main paper due to space constraints.

\subsection{Scale-invariant SNR (SI-SNR)}

For measuring the fidelity of the reconstruction of the on-screen component, we also report the scale-invariant signal-to-noise ratio (SI-SNR) \cite{LeRoux2018a} of the on-screen estimate $\hat{x}^\mathrm{on}$, which is defined as follows:
\begin{equation}
\begin{gathered}
    \text{SI-SNR}({x}, \hat{x}) = 10 \log_{10} 
    \frac
    {\| \alpha {x}\|^2}
    {\| \alpha {x} - \hat{x}\|^2},
    \enskip
    \alpha = \textrm{argmin}_{a} \| a {x} - \hat{x}\|^2 
    = \frac
        {{x}^T \hat{x}}
        {\|{x}\|^2}.
\end{gathered}
\end{equation}
In the course of our experiments, we discovered that SI-SNR is a potentially optimistic measure of on-screen separation quality, especially when comparing SI-SNR to OSR. Since SI-SNR scales the reference signal to compensate for gain errors on the estimate, this means that a model can predict scaled-down probabilities that maximize OSR without affecting SI-SNR and producing an on-screen estimate that has a gain error.
However, SI-SNR is useful in cases where one can obtain higher OSR values by scaling down the estimate and obtain falsely higher OSR numbers (see Table \ref{tab:sisnr_extra_results}).

\begin{table}[h]
    \centering
    \caption{Evaluation results, including median SI-SNR and SNR, along with the AUC-ROC,
    for filtered off-screen background (from \cite{tzinis2021into}) and unfiltered random background (our new proposed) test sets at 16 FPS.  For each model, calibration to 6 dB OSR is performed separately on the filtered and unfiltered test sets. ``AudioScope$^*$'' is our improved implementation of AudioScope \cite{tzinis2021into} and using our proposed training procedure while also pre-training the audio source separation module. 
    }
    \label{tab:sisnr_extra_results}
\scalebox{0.81}{
\begin{tabular}{
l
@{\hskip 5\tabcolsep}
c
@{\hskip 25\tabcolsep}
c @{\hskip 5\tabcolsep}
c @{\hskip 5\tabcolsep}
c
@{\hskip 10\tabcolsep}
c @{\hskip 5\tabcolsep}
c @{\hskip 5\tabcolsep}
c
}
\multicolumn{2}{c}{}
&\multicolumn{3}{c}{Filtered \cite{tzinis2021into}}
&\multicolumn{3}{c}{Unfiltered (new proposed)}\\
\cmidrule(r{\dimexpr 10\tabcolsep}){3-5} 
\cmidrule(r{\dimexpr 5\tabcolsep}){6-8}
\multicolumn{2}{l}{AV alignment}
&{\bf SI-SNR}
&{\bf SNR}
&{\bf AUC}
&{\bf SI-SNR}
&{\bf SNR}
&{\bf AUC}
\\
\midrule
\multicolumn{2}{l}{\multirow{1}{*}{No processing ($\hat{x}^\mathrm{on}=x$ with 0dB OSR)}}  &  4.4  & 4.4 & -- & 2.5  & 2.5 & -- \\
\multicolumn{2}{l}{\multirow{1}{*}{No processing ($\hat{x}^\mathrm{on}=x / 2$ with 6dB OSR)}}  &  4.4  & 4.7 & -- & 2.5  & 4.1 & -- \\
\midrule
\multicolumn{2}{l}{\multirow{1}{*}{AudioScope* (previous state-of-the-art)}} & 9.5 & 5.9 & 0.77 & 5.5 & 5.2 & 0.71 \\ \midrule
\multicolumn{2}{l}{\multirow{1}{*}{AudioScopeV2 with Joint CMA (Ours)}}  & \textbf{10.8} & \textbf{10.0} & \textbf{0.85} & {\bf 7.8} & {\bf 7.7} & {\bf 0.84} \\ 
\midrule
\end{tabular}}
\end{table}

From Table \ref{tab:sisnr_extra_results}, we see that by halving the gain of the input mixture itself, we are able to increase the SNR 2.5 dB $\rightarrow$ 4.1 dB, which is also close to the performance that the previous state-of-the-art model obtains when trained with our recipe on the unfiltered YFCC100M data (SNR of 5.2 dB). This is due to SNR providing an overly optimistic measure when the estimate's gain is less than the correct gain 
\cite{LeRoux2018a}, and this is the main reason that we also include the SI-SNR metric to all the reported ablation studies. 

On the contrary, reporting SI-SNR only could also sometimes lead to false conclusions. For instance, our highest performing model evaluated on the filtered YFCC100M data, joint CMA, clearly outperforms the previous state-of-the-art model by 4.1 (5.9 dB $\rightarrow$ 10.0 dB) in terms of SNR but only by 1.3 (9.5 dB $\rightarrow$ 10.8 dB) in terms of SI-SNR. This is due to the fact that a model is able to always perform higher OSR by scaling down all the estimated probabilities, at the expense of decreasing the gain on the on-screen estimate.
As a result, the SNR metric shows that our method is able to estimate a more accurate gain of the on-screen estimate and truly increase its reconstruction fidelity.

\subsection{Ablations for proposed separable SA}

The hyperparameters for our architecture were chosen according to informal tuning during model development. To determine if there are better settings of our proposed attention-based architectures, we performed a number of ablations. Since results were fairly similar among different architectures, we selected our proposed unsupervised separable SA model as a representative, and retrained this model with a number of different settings. The results are shown in Table \ref{tab:ablations}.

\begin{table*}[h]
    \centering
    \caption{Ablation results, including median SI-SNR, SNR, and OSR, along with the AUC-ROC, for the proposed 16 FPS unsupervised separable SA with 4 attention heads, embedding dimension $D=128$, $L=4$ blocks, $8\times8$ spatial locations from the visual embedding network, and a dropout rate of 0.2. We perform the ablations on the introduced unfiltered random background YFCC100M test partition at 2 calibration points with a specified OSR at 6 and 10 dB. The separation performance of the oracle MixIT* assignment is 10.4 dB SNR and 10.1 dB SI-SNR.}
\label{tab:ablations}
    \scalebox{0.99}{
    \begin{tabular}{
@{}l
@{\hskip 10\tabcolsep}r
@{\hskip 5\tabcolsep}r
@{\hskip 5\tabcolsep}r
@{\hskip 5\tabcolsep}r
@{\hskip 10\tabcolsep}r
@{\hskip 5\tabcolsep}r
@{\hskip 5\tabcolsep}r
@{\hskip 5\tabcolsep}r}
&\multicolumn{4}{c}{$\mathrm{OSR}_\mathrm{target}=6$dB}
&\multicolumn{4}{c}{$\mathrm{OSR}_\mathrm{target}=10$dB}
\\
\cmidrule(r{\dimexpr 10\tabcolsep}){2-5}
\cmidrule(r{\dimexpr 0\tabcolsep}){6-9}
Ablation
&{\bf SI-SNR}
&{\bf SNR}
&{\bf OSR}
&{\bf AUC}
&{\bf SI-SNR}
&{\bf SNR}
&{\bf OSR}
&{\bf AUC}
\\
-- & 7.3 & 6.6 & 6.0 & 0.80 & 6.6 & 5.5 & 10.0 & 0.80\\
\midrule
4x4 spatial & 8.4 & 7.0 & 6.0 & 0.83 & 7.8 & 5.8 & 10.0 & 0.83\\
No dropout & 6.9 & 7.0 & 6.0 & 0.81 & 7.0 & 6.1 & 10.0 & 0.81\\
2 heads & 8.1 & 7.4 & 6.0 & 0.83 & 8.1 & 6.3 & 10.0 & 0.83\\
8 heads & 7.2 & 6.2 & 6.0 & 0.77 & 6.7 & 4.7 & 10.0 & 0.77\\
2 blocks & 7.1 & 6.5 & 6.0 & 0.80 & 7.2 & 5.3 & 10.0 & 0.80\\
8 blocks & 2.5 & 3.5 & 1.1 & 0.48 & 2.5 & 3.5 & 1.1 & 0.48\\
D=256 & 7.6 & 6.8 & 6.0 & 0.78 & 8.2 & 5.7 & 10.0 & 0.78\\
D=64 & 7.4 & 7.2 & 6.0 & 0.80 & 7.5 & 6.0 & 10.0 & 0.80\\
\bottomrule
\end{tabular}}
\end{table*}

Though there are some settings that work a bit better for separable SA in these ablations (e.g.\ using fewer attention heads), we decided to present results with the original default settings in our main paper.
Generally these results indicate that some additional performance may be available with additional hyperparameter exploration. We do observe that a deeper model with 8 blocks seems to have a negative effect on the performance across the board, perhaps because of overfitting to the training set, or failure to converge.   

\subsection{Unsupervised vs semi-supervised results}
We also experimented with semi-supervised training, where we leverage videos annotated for the presence of on-screen and off-screen sounds.
Semi-supervised training uses unsupervised NOn examples as described in the main paper, plus additional ``human-labeled on-screen-only (LOn)'' and ``human-labeled off-screen-only (LOff)'' examples. For these examples we use both single-mixture and MoM versions. LOn single-mixture examples are just video frames and audio drawn from a unanimously-rated on-screen-only video. LOn MoM examples are the same as LOn single-mixture, except that synthetic off-screen audio from a random video is added. LOff single-mixture and MoM examples are the same as LOn examples, except that unanimously-rated off-screen-only videos are used for the primary video frames and audio. 
An exact cross-entropy loss
is used for training the on-screen classifier with these labeled examples, where the MixIT assignments are used as classifier labels $y$ for on-screen examples, and the classifier labels $y$ are set to zero for off-screen examples.
For semi-supervised training, examples in the batch are dynamically sampled, with 50\% NOn MoM, 12.5\% LOn single-mixture, 12.5\% LOn MoM, 12.5\% LOff single-mixture, and 12.5\% LOff MoM.

\begin{table}[h]
    \centering
    \caption{Evaluation results, including median SI-SNR and SNR, along with the AUC-ROC,  on the unfiltered random background test set at 16 FPS for the proposed models for unsupervised or semi-supervised training.
    Models are calibrated to 2 different OSR levels: 6 dB and 10 dB. 
    The detection capability of the models for the on-screen presence of the estimated sources, measured by the weighted area under the ROC curve (AUC), remains unaltered with different specified OSR operating points.}
    \label{tab:semisup_unsup}
\scalebox{0.99}{
\begin{tabular}{l @{\hskip 5\tabcolsep} l
@{\hskip 5\tabcolsep} c 
@{\hskip 10\tabcolsep} cc 
@{\hskip 10\tabcolsep} cc
@{\hskip 15\tabcolsep} c}
\multicolumn{2}{c}{} & Semi- &
\multicolumn{2}{l}{$\mathrm{OSR}_\mathrm{target}=6$dB}  & \multicolumn{2}{c}{$\mathrm{OSR}_\mathrm{target}=10$dB} & \\
\cmidrule(r{\dimexpr 10\tabcolsep}){4-5} 
\cmidrule(r{\dimexpr 15\tabcolsep}){6-7} 
\multicolumn{2}{l}{AV alignment}
& \phantom{   } Supervised \phantom{   }
&{\bf SI-SNR}
&{\bf SNR}
&{\bf SI-SNR}
&{\bf SNR}
&{\bf AUC}
\\
\midrule
\multicolumn{1}{l}{\multirow{4}{*}{SA}} & \multicolumn{1}{l}{\multirow{2}{*}{Joint}} & & 7.7 & 7.7 & 7.6 & 6.9 & 0.83 \\
& & \checkmark & 6.5 & 5.0 & 6.5 & 5.5 & 0.80 \\
 \cline{2-8}
 & \multicolumn{1}{l}{\multirow{2}{*}{Sep.}} & & 7.3 & 6.6 & 6.6 & 5.5 & 0.80 \\
& & \checkmark & 7.2 & 7.0 & 8.0 & 6.4 & 0.88 \\
  \cline{1-8}
 \multicolumn{1}{l}{\multirow{4}{*}{CMA}} & \multicolumn{1}{l}{\multirow{2}{*}{Joint}} & & \textbf{7.8} & 7.7 & \textbf{8.3} & 6.7 & 0.84 \\
& & \checkmark & 6.2 & 6.5 & 6.4 & 5.6 & 0.83\\
 \cline{2-8}
 & \multicolumn{1}{l}{\multirow{2}{*}{Sep.}} & & 7.5 & 7.1 & 7.2 & 5.8 & 0.80 \\
& & \checkmark & 6.3 & 6.5 & 7.4 & 6.7 & {\bf 0.89} \\
\midrule
\end{tabular}}
\end{table}

We show the results using our proposed models at 16 FPS and at 2 different OSR target levels in Table \ref{tab:semisup_unsup}. Our results are somewhat mixed. %
For separable models, AUC-ROC improves by up to 0.09, but does not improve for joint models.
Also, even when AUC-ROC improves, this does generally improve calibrated on-screen reconstruction in terms of SI-SNR and SNR.
We think this is due to the fact that AUC-ROC is measured across all the possible operating points of the classifier, whereas the SNR evaluation can only be obtained after choosing a specified OSR target level. We postulate that one might be able to extend our work using appropriate signal-level reconstruction losses and also improve reconstruction fidelity performance for semi-supervised cases, but we defer this to future work.

\subsection{Ablation on selected calibration points}
Another important aspect of our proposed calibration method is that the tolerance of off-screen sound interference can be specified. To that end, we show the importance of our method by comparing to uncalibrated model evaluations at a few interesting OSR operating points. The results of this ablation study are in Table \ref{tab:abl_calibaration_results}.

\begin{table}[h]
    \centering
    \caption{Evaluation results, including median SI-SNR, SNR, and OSR, along with the AUC-ROC, on the unfiltered random background test set at 16 FPS for the proposed models with different levels of calibrated OSR levels. All models have been trained using our proposed unsupervised learning procedure on the unfiltered dataset YFCC100M train partition while also pre-training the audio source separation network. We show the results when we use the raw pre-trained models with no calibration and when we calibrate AudioScopeV2 at 3 different OSR levels of 6, 10, and 15 dB. The detection capability of the models for the on-screen presence of the estimated sources, measured by the weighted area under the ROC curve (AUC), remains unaltered with different specified OSR operating points.}
    \label{tab:abl_calibaration_results}
\scalebox{0.99}{
\begin{tabular}{l @{\hskip 40\tabcolsep} c c  @{\hskip 15\tabcolsep}
c
@{\hskip 10\tabcolsep} c
@{\hskip 10\tabcolsep} c
@{\hskip 10\tabcolsep} c
}
Calibration &
\multicolumn{2}{c}{ AV alignment \phantom{yolo}}
&{\bf SI-SNR}
&{\bf SNR}
&{\bf OSR}
&{\bf AUC}
\\
\midrule
\multicolumn{1}{l}{\multirow{4}{*}{{Uncalibrated}}}    & \multicolumn{1}{l}{\multirow{2}{*}{SA}} & \multicolumn{1}{l}{\multirow{1}{*}{Joint}} & 7.8 & 7.1 & 8.3 & 0.83\\
& & \multicolumn{1}{l}{\multirow{1}{*}{Sep.}} & 6.7 & 5.3 & 10.8 & 0.80\\
    & \multicolumn{1}{l}{\multirow{2}{*}{CMA}} & \multicolumn{1}{l}{\multirow{1}{*}{Joint}} & \textbf{8.3} & 7.0 & 8.7 & \textbf{0.84}\\
    & & \multicolumn{1}{l}{\multirow{1}{*}{Sep.}} & 7.2 & 5.8 & 9.3 & 0.80\\
\midrule
\multicolumn{1}{l}{\multirow{4}{*}{{$\mathrm{OSR}_\mathrm{target}=6$dB}}}    & \multicolumn{1}{l}{\multirow{2}{*}{SA}} & \multicolumn{1}{l}{\multirow{1}{*}{Joint}} & 7.7 & \textbf{7.7} & \multirow{4}{*}{6.0} & 0.83\\
& & \multicolumn{1}{l}{\multirow{1}{*}{Sep.}} & 7.3 & 6.6 & & 0.80\\
    & \multicolumn{1}{l}{\multirow{2}{*}{CMA}} & \multicolumn{1}{l}{\multirow{1}{*}{Joint}} & 7.8 & \textbf{7.7} &  & \textbf{0.84}\\
    & & \multicolumn{1}{l}{\multirow{1}{*}{Sep.}} & 7.5 & 7.1 & & 0.80\\
\midrule
\multicolumn{1}{l}{\multirow{4}{*}{{$\mathrm{OSR}_\mathrm{target}=10$dB}}}    & \multicolumn{1}{l}{\multirow{2}{*}{SA}} & \multicolumn{1}{l}{\multirow{1}{*}{Joint}} & 7.6 & 6.9 & \multirow{4}{*}{10.0} & 0.83\\
& & \multicolumn{1}{l}{\multirow{1}{*}{Sep.}} & 6.6 & 5.5 & & 0.80\\
    & \multicolumn{1}{l}{\multirow{2}{*}{CMA}} & \multicolumn{1}{l}{\multirow{1}{*}{Joint}} & \textbf{8.3} & 6.7 &  & \textbf{0.84}\\
    & & \multicolumn{1}{l}{\multirow{1}{*}{Sep.}} & 7.2 & 5.8 &  & 0.80\\
\midrule
\multicolumn{1}{l}{\multirow{4}{*}{{$\mathrm{OSR}_\mathrm{target}=15$dB}}}    & \multicolumn{1}{l}{\multirow{2}{*}{SA}} & \multicolumn{1}{l}{\multirow{1}{*}{Joint}} & 7.6 & 6.0 & \multirow{4}{*}{15.0} & 0.83\\
& & \multicolumn{1}{l}{\multirow{1}{*}{Sep.}} & 7.0 & 3.4 &  & 0.80\\
    & \multicolumn{1}{l}{\multirow{2}{*}{CMA}} & \multicolumn{1}{l}{\multirow{1}{*}{Joint}} & \textbf{8.3} & 5.5 &  & \textbf{0.84}\\
    & & \multicolumn{1}{l}{\multirow{1}{*}{Sep.}} & 7.2 & 4.3 &  & 0.80\\
\midrule

\end{tabular}}
\end{table}

Notice that the uncalibrated joint SA seems to be performing significantly better than separable SA in terms of on-screen 
SNR: $7.1$ dB vs $5.3$ dB. However, if we also consider the ability of these models to suppress the off-screen component, the comparison becomes less clear since joint SA and separable SA obtain $8.3$ dB and $10.8$ dB OSR, respectively. To allow a more fair and easy-to-understand comparison of these models, our proposed calibration method can compensate for the OSR mismatch by tuning both models to specified OSR target level (e.g.\ $\mathrm{OSR}_\mathrm{target}=6$dB). After doing this, we can see that the actual SNR difference is almost $1$ dB (joint SA and separable SA obtain $7.7$ dB and $6.6$ dB SNR, respectively).

Unsurprisingly, for increasing levels of specified OSR target levels, the SNR performance on the reconstruction of the on-screen component gradually declines. However, we want to emphasize that in practice the operating point can be calibrated on validation data, according to the needs of an application, or according to user preferences. We also see that the separable versions of SA and CMA are able to perform on-par with the much more computationally expensive joint counterparts across all the specified OSR target levels.

\section{Evaluation on restricted-domain datasets}
\label{sec:eval}

Though our goal is to train a general-purpose on-screen separation model, it is interesting to see how well AudioScopeV2 performs on specialized domains without any further fine-tuning. 
We evaluated our models on these datasets, but were unable to include these results in the main paper due to space constraints. In the following subsections, we present these results.

Our purpose is not to compete with prior methods that use carefully curated training data to match these test sets (e.g.\ Mandarin has a very small corresponding training set, and approaches in the literature have crafted custom training sets) as we anticipate that our general model will do less well than a model specifically trained towards a more specialized task and domain.
We show these mismatched evaluations to examine whether more general approaches like ours could be used in handling more specialized tasks.

\subsection{MUSIC dataset for audio-visual musical instrument separation}

To measure performance on a non-speech task, we evaluated our proposed models on the MUSIC dataset \cite{zhao2018soundOfPixels}, which is a dataset of single-source videos of people playing musical instruments. We used the standard protocol \cite{gao2019co} to prepare the dataset, where we created 10 mixtures for each of the 55 possible pairs of 11 instrument classes, for a total of 550 examples. For each example, we use the video for one of the instruments as the video input to AudioScopeV2, and we do this for both videos (thus the total number of examples is $2\cdot550=1100$). Performance is measuring using \texttt{bss\_eval\_sources} \cite{vincent2006performance}, which measures signal-to-distortion ratio (SDR), signal-to-inference ratio (SIR), and signal-to-artifact ratio (SAR). These measures find an optimal time-invariant 512-tap filter that can be applied to the reference to maximize SDR. We compare our methods to a number of other recent approaches that also evaluate on this dataset.

\begin{table*}[h]
    \centering
    \caption{Results on the MUSIC dataset, including mean SDR, SIR, and SAR.}
    \label{tab:music}
\begin{tabular}{lcrrrr}
\toprule
                                                                  Model &
                                                                  Oracle?
                                                                  & Training data &            SDR &            SIR &            SAR \\
\midrule
                        Sound-of-Pixels \cite{zhao2018soundOfPixels} &         & MUSIC &            5.4 &           11.0 &            9.8 \\
                                  Sound-of-Motions \cite{zhao2019soundofmotions} &         & MUSIC &            4.8 &           11.0 &            8.7 \\
                                          MP-Net \cite{xu2019recursive} &         & MUSIC &            5.7 &           11.4 &           10.4 \\
                                         Co-Separation \cite{gao2019co} &         & MUSIC &            7.4 &           13.7 &           10.8 \\
 Cascaded Opponent Filter \cite{zhu2021visually_guided_SS_localization} &         & MUSIC &           10.1 &           16.7 &           13.0 \\
                      A(Res-50, att) + S(DV3P) \cite{zhu2020separating} &         & MUSIC &            9.4 &           15.6 &           12.7 \\
                   A(Res-50, class.) + S(DV3P) \cite{zhu2020separating} &         & MUSIC &           10.6 &           17.2 &           12.8 \\
                                      AVSGS \cite{chatterjee2021visual} &         & MUSIC &  \textbf{11.4} &  \textbf{17.3} &           13.5 \\
\midrule
                 AudioScope \cite{tzinis2021into} $\hat{x}^\mathrm{on}$ &      & Filtered YFCC100M &           -0.5 &            2.8 &           11.2 \\
                       AudioScope \cite{tzinis2021into} MixIT* & \checkmark &      Filtered YFCC100M &            8.8 &           13.0 &           13.1 \\
\midrule
                                               Joint SA (unsup), MixIT* & \checkmark &      YFCC100M &           10.0 &           14.1 &           14.6 \\
                                Joint SA (unsup), $\hat{x}^\mathrm{on}$ & &      YFCC100M &            2.1 &            3.6 &           18.1 \\
                                            Joint SA (semi-sup), MixIT* & \checkmark &      YFCC100M &            9.8 &           13.8 &           14.7 \\
                             Joint SA (semi-sup), $\hat{x}^\mathrm{on}$ & &      YFCC100M &            1.7 &            3.3 &           18.3 \\
                                               Sep.\ SA (unsup), MixIT* & \checkmark &      YFCC100M &           10.0 &           14.1 &           14.4 \\
                                Sep.\ SA (unsup), $\hat{x}^\mathrm{on}$ & &      YFCC100M &            0.4 &            1.3 &           19.7 \\
                                            Sep.\ SA (semi-sup), MixIT* & \checkmark &      YFCC100M &            9.8 &           13.8 &           14.6 \\
                             Sep.\ SA (semi-sup), $\hat{x}^\mathrm{on}$ & &      YFCC100M &            0.4 &            1.6 &           18.1 \\
                                              Joint CMA (unsup), MixIT* & \checkmark &      YFCC100M &            9.7 &           13.9 &           14.6 \\
                               Joint CMA (unsup), $\hat{x}^\mathrm{on}$ & &      YFCC100M &            3.1 &            4.8 &           18.6 \\
                                           Joint CMA (semi-sup), MixIT* & \checkmark &      YFCC100M &            9.8 &           13.8 &           14.5 \\
                            Joint CMA (semi-sup), $\hat{x}^\mathrm{on}$ & &      YFCC100M &            2.1 &            3.5 &           19.1 \\
                                              Sep.\ CMA (unsup), MixIT* & \checkmark &      YFCC100M &            9.9 &           13.9 &           14.5 \\
                               Sep.\ CMA (unsup), $\hat{x}^\mathrm{on}$ & &      YFCC100M &            1.9 &            3.3 &           17.5 \\
                                           Sep.\ CMA (semi-sup), MixIT* & \checkmark &      YFCC100M &            9.9 &           14.0 &           14.4 \\
                            Sep.\ CMA (semi-sup), $\hat{x}^\mathrm{on}$ & &      YFCC100M &           0.0 &            0.3 &  \textbf{25.4} \\
\bottomrule
\end{tabular}
    
\end{table*}

There are a few things to notice from these results. First, neither our proposed methods nor AudioScope reach state-of-the-art performance on this task compared to models trained on matched data with various degrees of supervision. However, the oracle MixIT* performance of the audio-only separation component of our proposed models, where separated sources are assigned to one of the ground-truth reference audio sources, is quite strong, within only 1.4 dB SDR for the best matched-training model (10.0 dB versus 11.4 dB). This MixIT* performance is also better than AudioScope, which only achieves 8.8 dB SDR. This improvement over AudioScope is presumably due to audio-only pre-training of our separation model on unfiltered YFCC100M with MixIT.

Unfortunately, the non-oracle $\hat{x}^\mathrm{on}$ output, which uses the predicted on-screen probabilities as mixing weights, still lags behind the oracle MixIT* scores. However, our proposed $\hat{x}^\mathrm{on}$ models do achieve better performance than AudioScope: in the best case, unsupervised joint CMA achieves 3.1 dB SDR, compared to -0.5 dB SDR for AudioScope, which is a significant boost. This suggests that using a wider variety of YFCC100M data helps AudioScopeV2 generalize, but that there may still be mismatch between videos in the MUSIC dataset and videos in YFCC100M. 
We anticipate that fine-tuning one of our proposed models on data from a target domain could help reduce this gap in performance.

\clearpage
\subsection{Mandarin dataset for audio-visual speech enhancement}

Audio-visual speech enhancement, which is the task of separating speech of an on-screen talker given video and noisy speech audio, has been explored by many recent works. To measure performance on this more restricted task, we use the Mandarin dataset \cite{hou2018audio}, which consists of video clips of 3-5 seconds of a person speaking Mandarin sentences, where background noise has been artificially added. The results are shown in Table \ref{tab:mandarin}, which includes several models from recent works which were trained specifically for the audio-visual speech enhancement task.

\begin{table*}[h]
    \centering
    \caption{Results  on Mandarin audio-visual speech enhancement evaluation set, in terms of mean SDR.}
    \label{tab:mandarin}
    \begin{tabular}{lcrr}
\toprule
                                                                          Model & Oracle? &                Training task &            SDR \\
\midrule
                                                 Hou et al. \cite{hou2018audio} & &              AV speech enhancement &            2.8 \\
                                         Ephrat et al. \cite{ephrat2018looking} & &             AV speech enhancement &            6.1 \\
 Gao and Grauman \cite{gao2021visualvoiceSpeechSeparationCrossModalConsistency} & &              AV speech enhancement &            6.7 \\
\midrule
                         AudioScope \cite{tzinis2021into} $\hat{x}^\mathrm{on}$ & &  AV universal on-screen sep. &            2.5 \\
                               AudioScope \cite{tzinis2021into} MixIT* & \checkmark &  AV universal on-screen sep. &            3.4 \\
\midrule
                               
                                                       Joint SA (unsup), MixIT*  & \checkmark &  Audio-only universal on-screen sep. &            9.8 \\
                                        Joint SA (unsup), $\hat{x}^\mathrm{on}$ & &  AV universal on-screen sep. &            1.5 \\
                                                    Joint SA (semi-sup), MixIT*  & \checkmark &  Audio-only universal on-screen sep. &            9.8 \\
                                     Joint SA (semi-sup), $\hat{x}^\mathrm{on}$ & &  AV universal on-screen sep. &           -0.2 \\
                                                       Sep.\ SA (unsup), MixIT*  & \checkmark &  Audio-only universal on-screen sep. &            9.6 \\
                                        Sep.\ SA (unsup), $\hat{x}^\mathrm{on}$ & &  AV universal on-screen sep. &           -0.1 \\
                                                    Sep.\ SA (semi-sup), MixIT*  & \checkmark &  Audio-only universal on-screen sep. &           10.0 \\
                                     Sep.\ SA (semi-sup), $\hat{x}^\mathrm{on}$ & &  AV universal on-screen sep. &            0.6 \\
                                                      Joint CMA (unsup), MixIT*  & \checkmark &  Audio-only universal on-screen sep. &            9.6 \\
                                       Joint CMA (unsup), $\hat{x}^\mathrm{on}$ & &  AV universal on-screen sep. &            2.3 \\
                                                   Joint CMA (semi-sup), MixIT*  & \checkmark &  Audio-only universal on-screen sep. &  \textbf{10.0} \\
                                    Joint CMA (semi-sup), $\hat{x}^\mathrm{on}$ & &  AV universal on-screen sep. &           -0.5 \\
                                                      Sep.\ CMA (unsup), MixIT*  & \checkmark &  Audio-only universal on-screen sep. &            9.7 \\
                                       Sep.\ CMA (unsup), $\hat{x}^\mathrm{on}$ & &  AV universal on-screen sep. &            1.8 \\
                                                   Sep.\ CMA (semi-sup), MixIT*  & \checkmark &  Audio-only universal on-screen sep. &            9.9 \\
                                    Sep.\ CMA (semi-sup), $\hat{x}^\mathrm{on}$ & &  AV universal on-screen sep. &            1.1 \\
\bottomrule
\end{tabular}
\end{table*}

First, note that our oracle MixIT* models outperform state-of-the-art on this dataset (10.0 dB SDR, versus 6.7 dB SDR for the best matched-training baseline). This implies that the general-purpose separation model pre-trained with MixIT is quite strong. Also, pre-training on unfiltered YFCC100M is perhaps why the MixIT* performance of our models is so much better than the MixIT* performance of AudioScope, which was only trained on filtered YFCC100M.

However, as with the MUSIC dataset, the non-oracle output of $\hat{x}^\mathrm{on}$ degrades compared to the oracle performance. We postulate that due to the mismatch between the Mandarin video data and YFCC100M video data, our proposed models score 2.3 dB in terms of SDR which could be potentially boosted using fine-tuning on matched video samples.

\section{Visualizations of attention maps}
\label{sec:visual}

\begin{figure}[htb!]
  \centering
  \includegraphics[width=1.\linewidth]{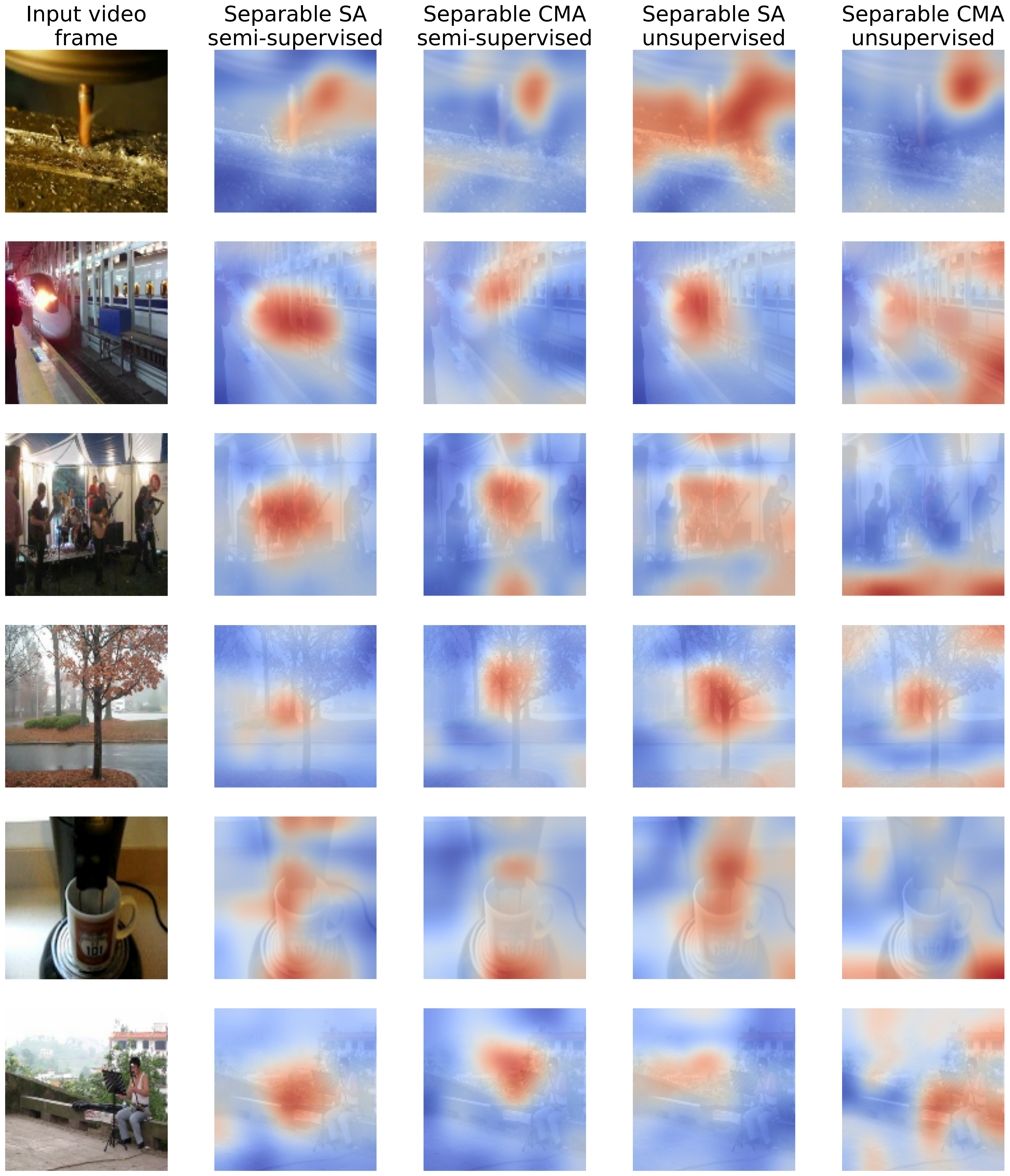}
  \caption{Weighted attention maps for different training conditions with the proposed audio-visual attention models using uncalibrated separable self-attention (SA) and cross-modal attention (CMA) for a given input video frame. The attention maps have been weighted using the on-screen estimated probability per corresponding source.
  }
  \label{fig:attention_maps_onscreen}
\end{figure}

In Figure \ref{fig:attention_maps_onscreen} several attention maps are displayed for the proposed audio-visual attention architectures. Each heatmap is derived by using the per-source on-screen classifier probability score to weight the attention map for the corresponding input frame while also summing across the different heads. In most of the displayed examples, the warm color regions co-locate with regions of the video frame that represents on-screen objects. Note that for the unsupervised models (two columns on the left), the attention maps are more dispersed across the image and sometimes focused on the background of each image (see rows 2-4 at the last column). We postulate that with strongly labeled data and/or pre-trained segmentation models one could possibly sharpen the accuracy of such attention mechanisms, but we defer that to future work.

\begin{figure}[htb!]
  \centering
  \includegraphics[width=1.\linewidth]{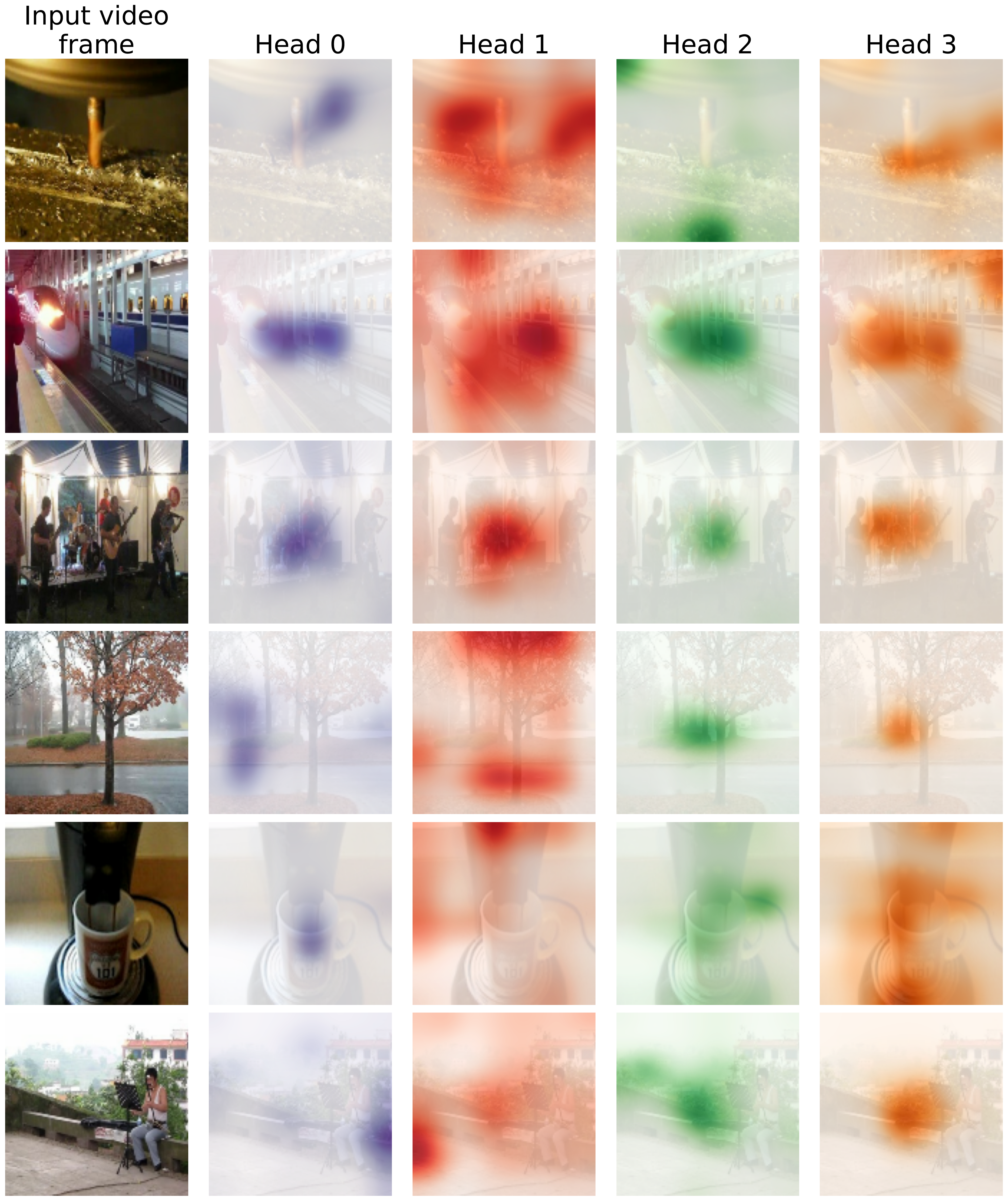}
  \caption{Attention maps obtained from the different heads of the first separable audio-visual self-attention (SA) layer.}
  \label{fig:heads_SA}
\end{figure}

\begin{figure}[htb!]
  \centering
  \includegraphics[width=1.\linewidth]{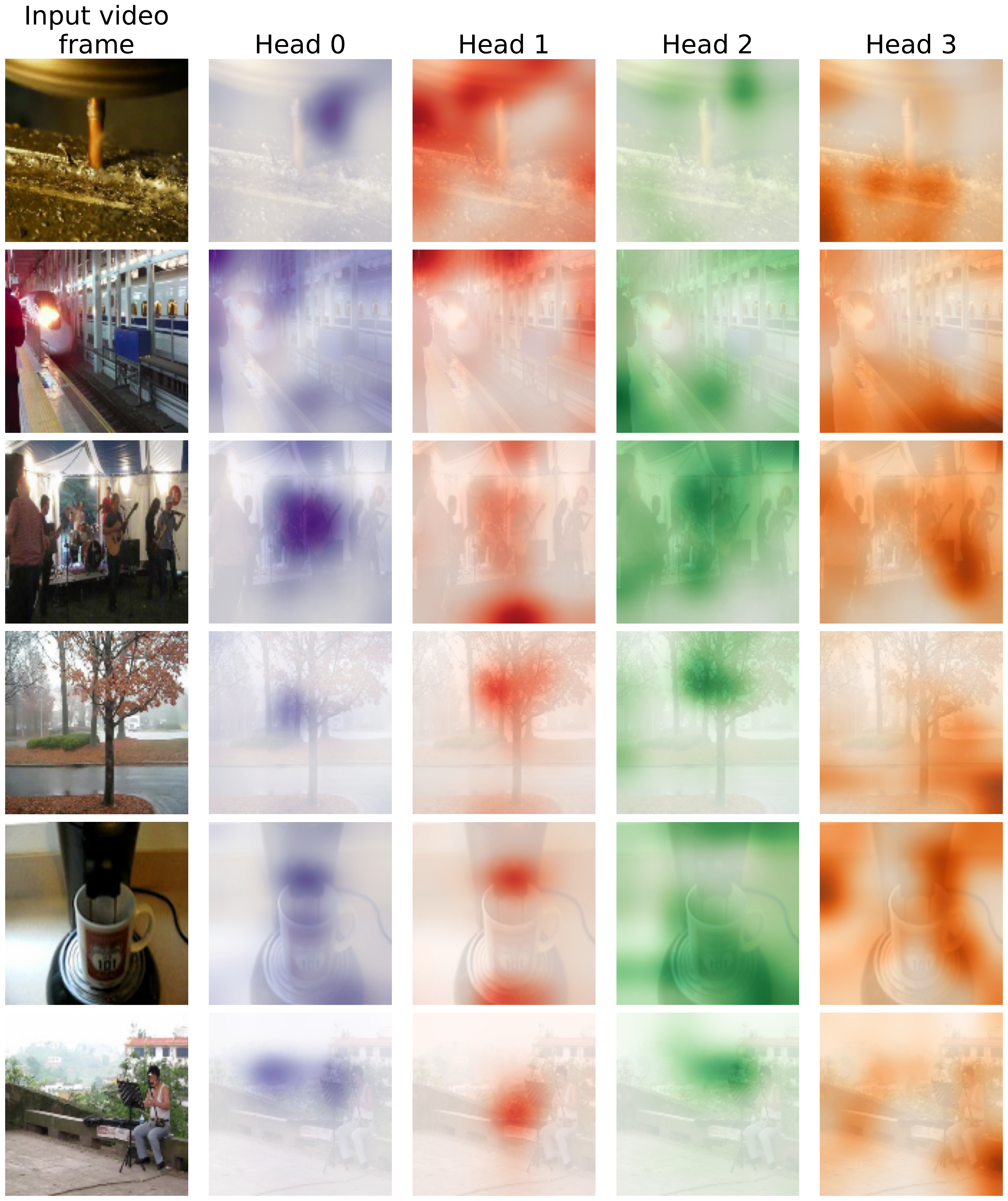}
  \caption{Attention maps obtained from the different heads of the first separable audio-visual cross-modal attention (CMA) layer.}
  \label{fig:heads_CMA}
\end{figure}

Although the analysis of the attention maps for the on-screen objects hints at the capability of the model to understand the audio-visual alignment, we would like to better understand the implicit representation obtained from the SA and CMA models. To help with this, several attention maps obtained by different heads of the first layer of the semi-supervised separable SA model and the semi-supervised separable CMA model are shown in Figures \ref{fig:heads_SA} and \ref{fig:heads_CMA}, respectively. Notice that in both cases of attention architectures there are heads which attend to different parts of the input frame potentially showing the expressiveness of the proposed layers even in the case of the more efficient separable variation. Interestingly, ``Head 1'' (third column) in Figure \ref{fig:heads_SA} seems to have a more disperse attention pattern, possibly showing that this head learns to attend more to the background compared to other heads e.g.\ ``Head 0''.

\clearpage

\section*{Attributions}
\noindent Still images from the following videos are used in Figures \ref{fig:attention_maps_onscreen}, \ref{fig:heads_SA}, and \ref{fig:heads_CMA}:

\noindent \href{https://multimedia-commons.s3-us-west-2.amazonaws.com/data/videos/mp4/18a/6e3/18a6e3f056add1e14a19c265f88d48f.mp4}{``Tiny End Mill'' by ukweli}, license: \href{http://creativecommons.org/licenses/by-sa/2.0/}{CC-BY-SA 2.0}

\noindent \href{https://multimedia-commons.s3-us-west-2.amazonaws.com/data/videos/mp4/1dc/55e/1dc55e1946d57a3c9eaf4367d73be318.mp4}{``Shinkansen'' by pauldesu.com}, license: \href{http://creativecommons.org/licenses/by/2.0/}{CC-BY 2.0}

\noindent \href{https://multimedia-commons.s3-us-west-2.amazonaws.com/data/videos/mp4/18f/259/18f259d0b6569a34e44f6728c7c9e1.mp4}{``A Sample of Broken Mouth Annie'' by Tobyotter}, license: \href{http://creativecommons.org/licenses/by/2.0/}{CC-BY 2.0}

\noindent \href{https://multimedia-commons.s3-us-west-2.amazonaws.com/data/videos/mp4/1c0/daa/1c0daac4d679c652f95aa26636496.mp4}{``A rainy autumn afternoon in Norcross'' by sylvar}, license: \href{http://creativecommons.org/licenses/by/2.0/}{CC-BY 2.0}

\noindent \href{https://multimedia-commons.s3-us-west-2.amazonaws.com/data/videos/mp4/131/df6/131df6d78bd55af3455a4cc9c8d55f2.mp4}{``3rd frothy Cup of java, good morning.'' by miheco}, license: \href{http://creativecommons.org/licenses/by-sa/2.0/}{CC-BY-SA 2.0}

\noindent \href{https://multimedia-commons.s3-us-west-2.amazonaws.com/data/videos/mp4/217/b82/217b82d79d61338eea38ef7a4fbfc85.mp4}{``MVI\_6134'' by kenner116}, license: \href{http://creativecommons.org/licenses/by/2.0/}{CC-BY 2.0}

\end{document}